%% file: main.tex
\documentclass{article}

\PassOptionsToPackage{numbers, compress}{natbib}
\usepackage[preprint]{neurips_2024}

\usepackage[utf8]{inputenc} % allow utf-8 input
\usepackage[T1]{fontenc}    % use 8-bit T1 fonts
\usepackage{hyperref}       % hyperlinks
\usepackage{url}            % simple URL typesetting
\usepackage{booktabs}       % professional-quality tables
\usepackage{amsfonts}       % blackboard math symbols
\usepackage{nicefrac}       % compact symbols for 1/2, etc.
\usepackage{microtype}      % microtypography
\usepackage{xcolor}         % colors

\usepackage{graphicx}
\usepackage{wrapfig}
\usepackage{algorithm}
\usepackage{algorithmic}
\usepackage{subcaption}

\usepackage{amsmath}
\usepackage{amssymb}
\usepackage{mathtools}
\usepackage{amsthm}
\usepackage{wasysym}
\usepackage{enumitem}
\usepackage{multirow}
\usepackage{adjustbox}
\usepackage{xspace}
\usepackage{cleveref}
\newcommand{\ie}{{\it i.e.}\xspace}
\newcommand{\eg}{{\it e.g.}\xspace}

\newcommand{\mtheta}{{\boldsymbol{\theta}}}

\title{Improving LLMs for Recommendation \\
with Out-Of-Vocabulary Tokens}

\author{
  Ting-Ji Huang$^{1,2}$ \ \ \ Jia-Qi Yang$^{1,2}$ \ \ \ Chunxu Shen$^3$ \ \ \ Kai-Qi Liu$^4$  \ \ \ \\ \textbf{De-Chuan Zhan}$^{1,2}$ \ \ \  \textbf{Han-Jia Ye}$^{1,2}$\thanks{Corresponding author, email: yehj@lamda.nju.edu.cn.}\\
  $^1$School of Artificial Intelligence, Nanjing University\\
  $^2$National Key Laboratory for Novel Software Technology, Nanjing University\\ $^3$WeChat Technical Architecture Department, Tencent Inc.\\ $^4$Software Institute, Nanjing University \\
}

\begin{document}

\maketitle

\input{abstract}

\input{intro}

\input{related}

\input{pre}

\input{method}

\input{experiments}

\input{conclusion}

% \newpage

{
\small
\bibliography{main}
\bibliographystyle{plainnat}
}

\input{appendix}

\end{document}

%% file: abstract.tex
\begin{abstract}

Characterizing users and items through vector representations is crucial for various tasks in recommender systems. Recent approaches attempt to apply Large Language Models (LLMs) in recommendation through a question\&answer format, where real users and items (\eg, Item No.2024) are represented with in-vocabulary tokens (\eg, ``item'', ``20'', ``24''). However, since LLMs are typically pretrained on natural language tasks, these in-vocabulary tokens lack the expressive power for distinctive users and items, thereby weakening the recommendation ability even after fine-tuning on recommendation tasks. In this paper, we explore how to effectively tokenize users and items in LLM-based recommender systems. We emphasize the role of out-of-vocabulary (OOV) tokens in addition to the in-vocabulary ones and claim the \textit{memorization} of OOV tokens that capture correlations of users/items as well as \textit{diversity} of OOV tokens. By clustering the learned representations from historical user-item interactions, we make the representations of user/item combinations share the same OOV tokens if they have similar properties. Furthermore, integrating these OOV tokens into the LLM’s vocabulary allows for better distinction between users and items and enhanced capture of user-item relationships during fine-tuning on downstream tasks. Our proposed framework outperforms existing state-of-the-art methods across various downstream recommendation tasks.

\end{abstract}

%% file: intro.tex
\section{Introduction}

\label{Section1}
Modern recommender systems (RS) play a crucial role in various applications like video recommendation~\cite{james2010}, e-commerce~\cite{qiwei2019}, and social networking~\cite{wenqi2019}. The recent advent of large language models (LLMs) offers a new direction of exploration in this realm. Models such as T5~\cite{colin2020exploring} and LLaMA~\cite{hugo2023llama}, training on massive natural language data, achieve impressive language understanding capabilities in text generation and conversation tasks~\cite{wayne2023a}. Their success drives explorations to use pre-trained LLMs as model backbones for handling various recommendation tasks like sequential recommendation, rating, and explanation task~\cite{qiwei2022,chuhan2021}. In such frameworks, we describe task inputs as a query and expect an LLM to directly answer task labels~\cite{zeyu2022, GengRecommendation22}. After fine-tuning through this Q\&A format, we could ask the LLM to recommend something for a user and receive a real item answered.

\begin{figure}[t!]
    \centering
    \subfloat[Characterizing items through tokens.\label{fig:teaser_a}]{
        \includegraphics[width=0.48\linewidth]{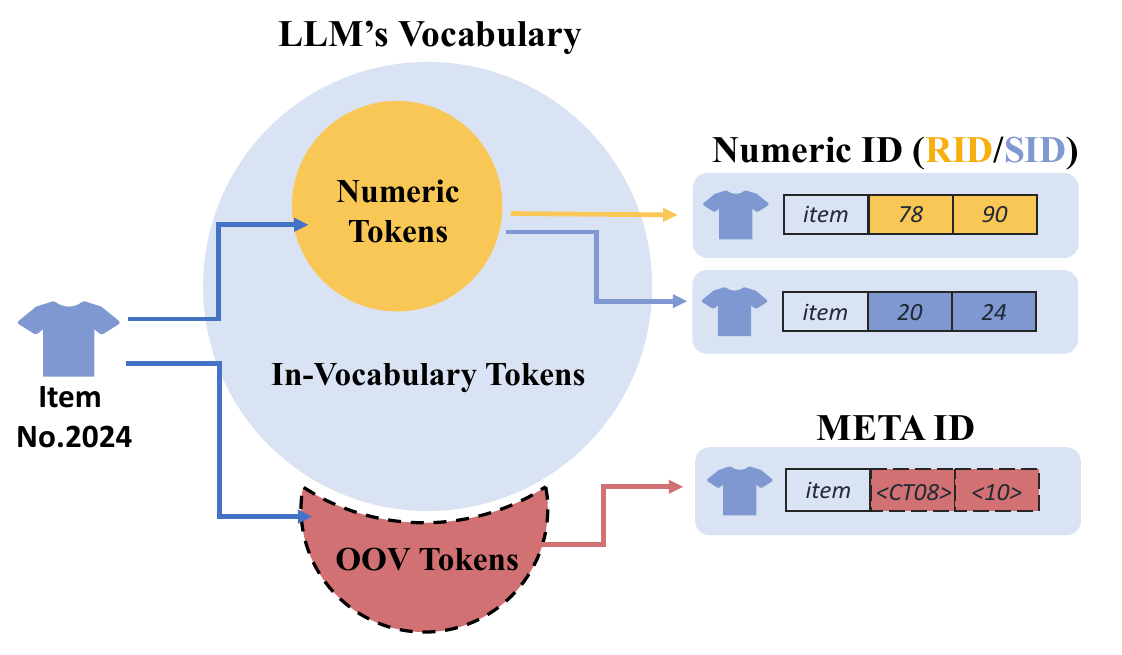}
    }
    \subfloat[Item similarity heatmaps with different IDs.\label{fig:teaser_b}]{
        \includegraphics[width=0.48\linewidth]{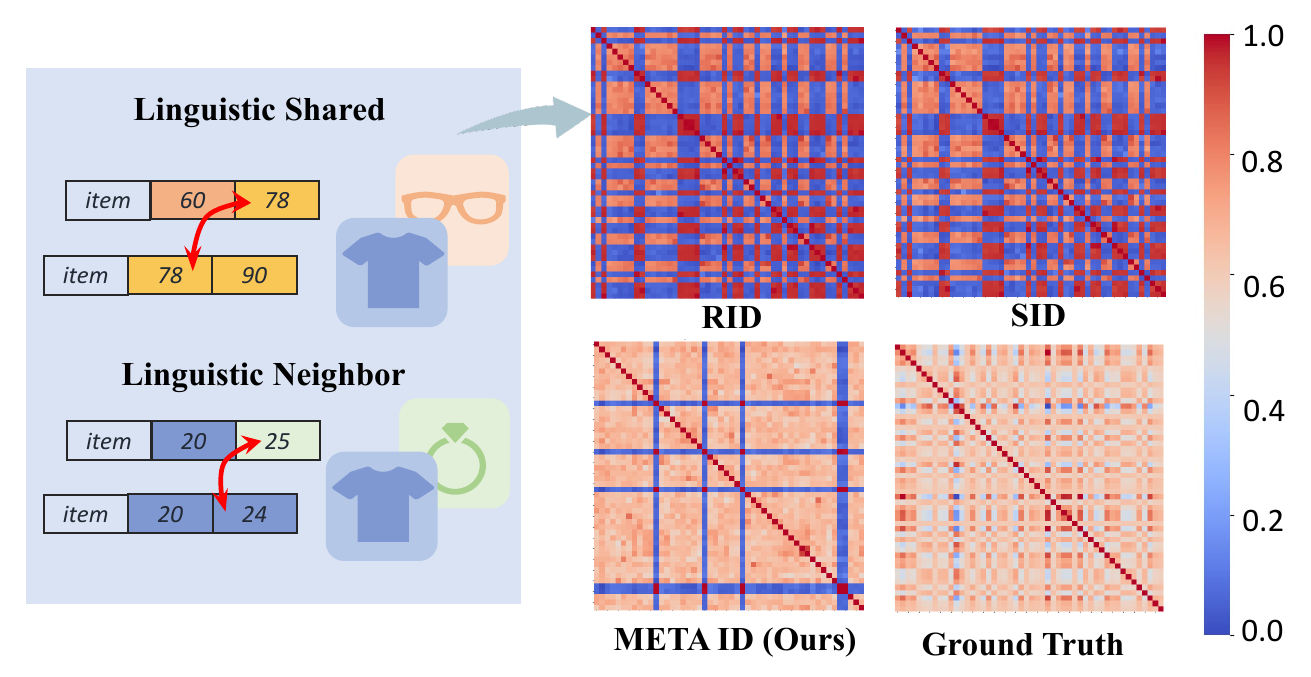}
    }
  \caption{ \small (a) Two ID strategies for characterizing items through tokens in LLM-based recommendation. (b) Left: Mapping problems caused by numeric tokens when characterizing items. Two unrelated items might share the same numeric tokens, and consecutive numeric tokens are treated similarly by LLMs. Right: Heatmaps of item representations similarity learned with Random ID (RID), Sequential ID (SID) and META ID on Toys dataset. We randomly sample 50 items and take their ID representations to calculate cosine similarity. The ground Truth is their adjusted cosine similarities in the Equation~\ref{eq:adjusted_sim}.}
  \label{fig:teaser}

\end{figure}

How to characterize users and items so that LLMs can learn their personalized information during fine-tuning? Since tokens are the basic building blocks for LLMs and represent the smallest unit of text the model can understand and process, we should allocate some tokens to represent users/items and let LLMs learn their characteristics through tokens. While early works~\cite{zeyu2022} directly represent a real item by text tokens (\eg, ``blue'', ``casual'', ``T-shirt''), LLMs might answer text that does not correspond to a real existing item~\cite{wenyue2023, ji2023survey}. Recent advancements explore representing users/items with \textit{Identifiers} (\textit{IDs})~\cite{GengRecommendation22,wei2022,yunfan2023,keqin2023}. As illustrated in Figure \ref{fig:teaser_a}, a real item (Item No.2024) can be indexed with a sequence of in-vocabulary tokens (``item'', ``20'', ``24''), and we call such a token combination as an ID, where a numeric ID indicates that we use numeric tokens (``20'', ``24'').

Characterizing users and items with numeric IDs is straightforward but also introduces the \textit{mapping problem}, where it is hard to align limited numeric tokens to thousands of items in recommender systems, and token combinations are prone to lead to language conflicts~\cite{keqin2023, wenyue2023}. As shown in Figure \ref{fig:teaser_b}, using numeric tokens to represent items leads to similar representations of distinctive items. One solution is to create OOV tokens specifically allocated for IDs to allow distinction, as in the previous work on Collaborative ID (CID)~\cite{wenyue2023}. 
However, our experiments indicate that the use of OOV tokens defined in CID may improve diversity, it is hard for LLM to learn the correlations of users/items during fine-tuning (see Section~\ref{Section3-3}). Therefore, we should utilize redefined OOV tokens capable of capturing user/item correlations (\textit{memorization}) while also distinguishing different items (\textit{diversity}), two dimensions that are positively correlated for the performance of recommendation tasks.

In this paper, we present META ID (META-path-guided IDentifier), a framework for characterizing users/items using out-of-vocabulary (OOV) tokens for LLM-based recommendations. Initially, we generate meta-paths to represent user-item interactions and then obtain user and item representations from a skip-gram model trained on these meta-paths. A meta-path is a sequence that represents interactions between users and items in a graph structure. By clustering these meta-path-based representations, we create hierarchical groups that serve as OOV tokens for constructing user and item IDs. This approach extends beyond previous research, which has predominantly focused on item IDs~\cite{GengRecommendation22, shashank2023recommender}, by making the representations of users/item IDs share the same OOV tokens if they have similar properties. Finally, integrating these OOV tokens into the LLM's vocabulary allows for better diversity and enhanced memorization during fine-tuning on downstream recommendation tasks. Additionally, we align the token embedding layer of LLMs with a linear transformation layer to enrich the OOV token representations as an augmentation. Our contributions are as follows:

\begin{itemize}[noitemsep,topsep=0pt,leftmargin=*]
\item We introduce memorization and diversity scores to evaluate ID representations in LLM-based recommender systems, focusing on capturing user/item correlations and ensuring diversity.
\item We develop META ID, which uses out-of-vocabulary tokens to characterize users/items, enhancing the memorization and diversity of their representations for LLMs.
\item The experiments show that META ID improves memorization and diversity score, which leads to improvements in various recommendation tasks.
\end{itemize}

%% file: related.tex
\section{Related Work}
\noindent{\bf Characterizing Items by IDs}. Modern recommendation models usually use unique IDs to represent users and items, which are subsequently converted to embedding vectors as learnable parameters~\cite{wenqi2023}. Common approaches include matrix factorization~\cite{yehuda2009matrix, steffen2009}, two-tower models~\cite{jinpeng2021cross} and deep neural networks~\cite{fei2019, wang2018, kun2020, yi2023model, xu2022contrastive, qi2023streaming}, which make predictions by examining historical user-item interactions to identify behavioral patterns and enable collaborative recommendations~\cite{yehuda2022advances}. Some recent approaches adopt the concept of ID to represent the token combinations that characterize items in LLMs~\cite{GengRecommendation22, wenyue2023}. We follow previous studies and also call these token combinations as ID. This paper aims to combine LLM with ID more efficiently by proposing a new ID construction method.

\noindent{\bf Instruction Tuning for Recommendation}. The integration of Large Language Models (LLMs) into diverse tasks has seen significant growth recently. Recent works fine-tuned pretrained language models on large-scale NLP datasets verbalized via human-readable prompts~\cite{victor2022multitask, jason2022finetuned}. These instruction tuning methods design prompts containing detailed task descriptions and adhere more to the natural language format. Driven by their exceptional natural language processing capabilities, researchers aim to transfer their linguistic ability to enhance recommender systems~\cite{lei2021,junjie2023}. These LLMs process user interactions as sequences of tokens and, through fine-tuning, predict users' future interests based on past activities~\cite{zhang2021language, zheng2023}. Moreover, some studies reframe tasks like retrieval, rating, and explanation generation as language comprehension tasks~\cite{zeyu2022, GengRecommendation22, jinming2023}, allowing LLMs to function as multi-task recommenders, producing recommendations and explanations with a unified architecture. In this paper, we apply LLMs based on instruction tuning of multi-task scenarios.

\noindent{\bf ID Construction with Tokens}. Recent studies explore using in-vocabulary tokens to characterize items in LLMs, where users and items are represented by token combinations called IDs. Early efforts such as P5~\cite{GengRecommendation22} convert user-item interactions into natural language formats using numeric IDs constructed of in-vocabulary tokens of the T5 model~\cite{colin2020exploring}. Further, sequential ID and collaborative ID are developed to enhance item information sharing~\cite{wenyue2023}. A relevant study~\cite{shashank2023recommender}, constructs IDs using hierarchical tokens through RQ-VAE. Despite these advancements in ID construction, challenges remain in the lack of ID construction criteria and the focus on item IDs. In contrast, our META ID approach assigns memorization and diversity scores for evaluation and constructs both user and item IDs, which improves the characterization of users and items.

%% file: pre.tex
\section{Preliminary}

We describe recommendation tasks (\eg, sequential recommendation) under the instruction tuning setting, in which all data such as user-item interactions are converted to natural language sequences in a question\&answer format (\eg, Input: \textit{Please recommend an item for user 2024 consider he has purchased item 2023}; Output: \textit{item 2024}). Details of this format are presented in Appendix~\ref{appendix3}.

\subsection{Instruction Tuning for Recommendation}
\label{Section3-1}
Given a user set $\mathbf{U}$, an item set $\mathbf{I}$, we formulate a recommendation task as a natural language instruction that pairs an input token sequence $\mathbf{x} = (x_1, ..., x_n, \boldsymbol{x}_u, \boldsymbol{x}_i)$ with a corresponding label token sequence $\mathbf{y} = (y_1, ..., y_m, \boldsymbol{x}_u, \boldsymbol{x}_i)$. The goal is to train and LLM $\mathcal{M}_{\mtheta}$ to generate $\mathbf{y}$ given $\mathbf{x}$. For simplification, we denote one token as $x_i$ or $y_i$, and represent the ID of an user $u \in \mathbf{U}$ by $\boldsymbol{x}_u = (x_{u_1}, x_{u_2}, ...)$ as its token combination, using the same representation format to a item ID $\boldsymbol{x}_i$.

The LLM $\mathcal{M}_{\mtheta}$ employs a token embedding layer $\boldsymbol{E}(\cdot)$ with parameters ${\mtheta}_{\boldsymbol{E}} \in {\mtheta}$, functioning as a lookup table that transforms each input token into a token embedding. It subsequently predicts the probability distribution of label tokens by forward propagation. The training objective is to minimize the negative log-likelihood of the label tokens conditioned on the input sequence and previously generated tokens, formulated as:
\begin{equation}
\small
\mtheta^* = \operatorname*{arg\,min}_{\mtheta} \mathcal{L}_{\mtheta} = -\sum_{j=1}^{|\mathbf{y}|} \log P_{\mtheta}\big(y_j \mid y_{<j}, \boldsymbol{E}(\mathbf{x})\big),
\end{equation}
where $\mtheta^*$ represents the optimal set of parameters that we aim to learn. This supervised learning approach helps the model to internalize personalized information about users and items through the token embeddings of their respective IDs, $\boldsymbol{x}_u$ and $\boldsymbol{x}_i$.

\subsection{Represent Users/Items with In-Vocabulary Tokens}
\label{Section3-2}

The token embedding layer $\boldsymbol{E}$ in LLMs transforms input tokens into corresponding token embeddings. Since users and items are also represented by IDs constructed of tokens, the corresponding token embeddings of ID $\boldsymbol{x}_u$ and $\boldsymbol{x}_i$ are the key to assess how LLMs capture and represent different items. 
We define the \textit{ID representation} for an ID $\boldsymbol{x}_i$ as the average embeddings of its token combinations:
\begin{equation}
\label{eq:3}
\mathbf{e}_i = \frac{1}{|\boldsymbol{x}_i|} \sum_{x \in \boldsymbol{x}_i} \boldsymbol{E}(x),
\end{equation}
where $\boldsymbol{E}(x)$ denotes the embedding vector of a token $x$. For example, an item identified by {``item~2024''} is represented by the average embedding vectors of three tokens (``item'', ``20'', ``24''). \begin{wrapfigure}{r}{0.4\textwidth}
    \centering
    \subfloat[T-SNE of ID representations.\label{fig:score-tsne}]{
        \includegraphics[width=5.5cm]{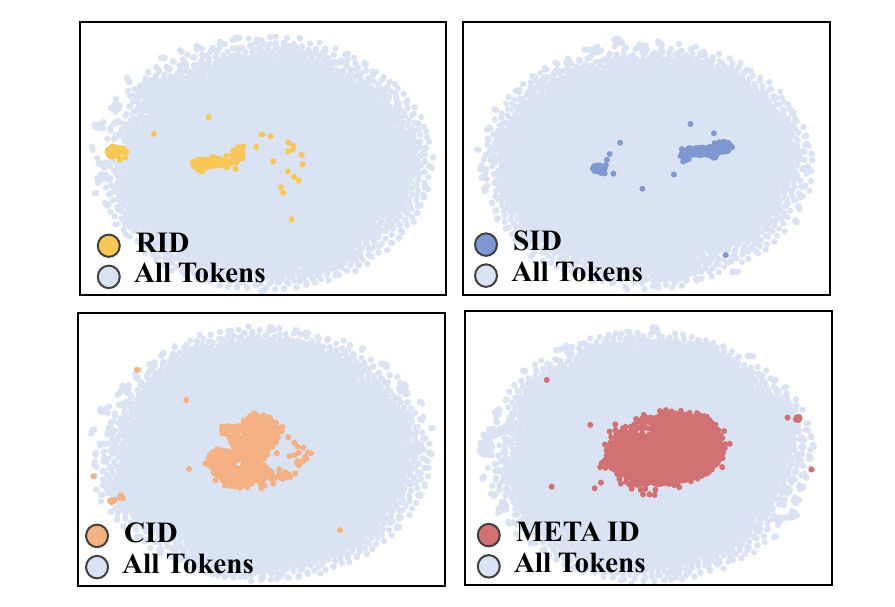}
    }
    \qquad
    \subfloat[Comparison of DS and MS.\label{fig:score}]{
        \includegraphics[width=5.3cm]{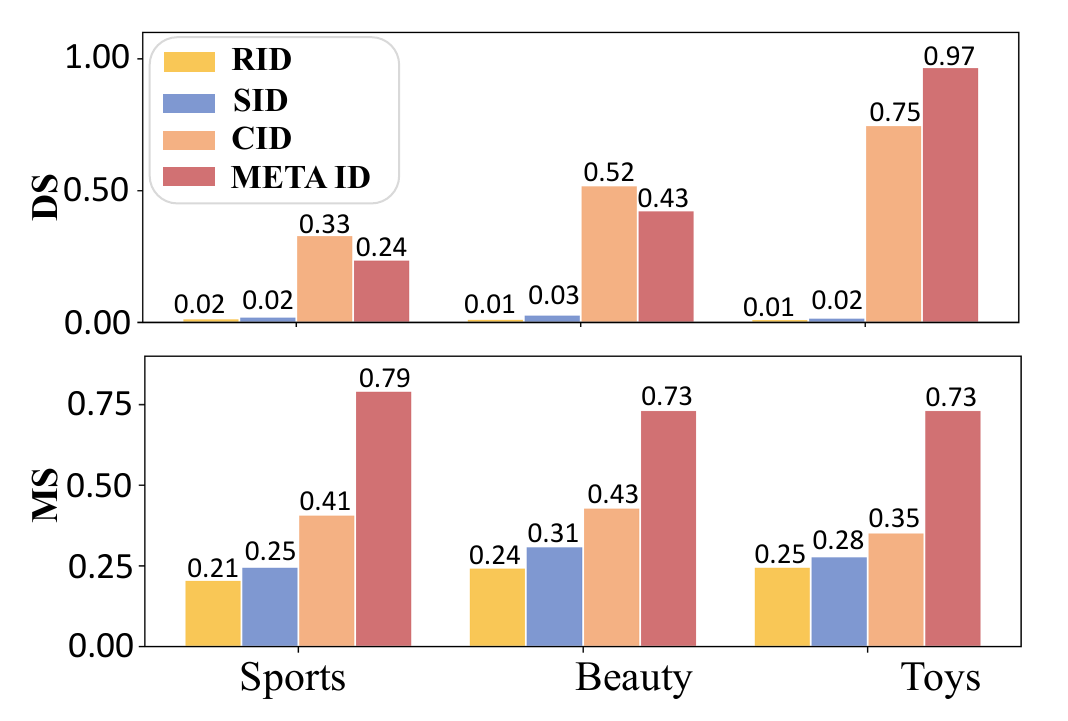}
    }
  \caption{ \small (a) T-SNE visualizations of ID representations on Sports dataset; (b) Comparison of DS and MS of ID representations from RID, SID, CID and META ID.}
\vspace{-40pt}
\label{fig:token}
\end{wrapfigure}

\subsection{Metric for Representation Evaluation}
\label{Section3-3}
As mentioned above, using numeric IDs to represent items leads to similar representations of distinctive items. For an intuitive explanation, we first visualize the cosine similarity matrix between items using heatmaps in Figure \ref{fig:teaser}, where RID and SID result in a large number of similar items due to semantic conflicts, and META ID constructed of OOV tokens shows distinguished similarity closer to the ground truth. We further plot the ID representations using T-SNE visualization in Figure~\ref{fig:score-tsne}. It is clear that RID and SID (using in-vocabulary tokens) shrink in a small place relative to CID (using OOV tokens), reflecting in-vocabulary tokens lacking expressive power for distinctive users and items. Details are in Section~\ref{Section6-3}. 

To quantify this, we introduce two metrics to assess the \textit{memorization} and \textit{diversity} of ID representations.

\noindent{\bf Diversity Score (DS)} is a metric designed to quantify the diversity of ID representations within LLMs. Items represented by ID representations should be easily distinguishable for the model. For a pair of items $i$ and $j$, their differentiation can be calculated using the Kullback-Leibler (KL) divergence on ID representations $\mathbf{e}_i \in E$ and $\mathbf{e}_j \in \mathbf{E}$ from Equation~\ref{eq:3}, which reflects how distinct they are in the model's embedding space: 
\begin{equation}
\small
\text{DS}(\mathbf{E}) = \frac{1}{2 N} \sum_{n=1}^{N}{\bigg[{D_{KL}(\mathbf{e}_i || \mathbf{e}_j) + D_{KL}(\mathbf{e}_j || \mathbf{e}_i)}\bigg]}.
\end{equation}
Here, $N$ represents the number of randomly selected item pairs, where we apply a sampling strategy to reduce the computational complexity, and we give a convergence analysis in Figure~\ref{fig:converge}.

\noindent{\textbf{Memorization Score (MS)}} quantifies the relationships captured by ID representations through measuring the similarity between items and users. We use the adjusted cosine similarity formula~\cite{badrul2001item}, to provides ground truth relational values between users and items, which are given by:
\begin{equation}
\small
\label{eq:adjusted_sim}
\text{sim}(i, j) = \frac{\sum_{u \in U} (R_{u,i} - \bar{R}_u) \cdot (R_{u,j} - \bar{R}_u)}{ {\sqrt{\sum_{u \in U}{(R_{u,i} - \bar{R}_u)^2} \cdot {\sum_{u \in U}{(R_{u,j} - \bar{R}_u)^2}}}}},
\end{equation}
where $R_{u,i}$ and $R_{u,j}$ denote user $u$'s ratings for items $i$ and $j$, and $\bar{R}_u$ is user $u$'s average ratings. To assess the relationship captured by learned ID representations, we employ Mean Square Error (MSE) to calculate the similarity bias on their cosine similarity with their corresponding ground truth relation values, which forms the basis of the MS:
\begin{equation}
\small
\text{MS}(\mathbf{E}) = \frac{2}{|N|(|N|-1)}\sum_{i,j \in N, i < j} \left( \frac{\mathbf{e}_i^\top \mathbf{e}_j}{|\mathbf{e}_i| |\mathbf{e}_j|} - \text{sim}(i, j)\right)^2.
\end{equation}

Our quantitative assessment of the quality of these ID representations using the introduced metrics reveals some interesting findings. In Figure~\ref{fig:score}, it shows that constructing IDs of in-vocabulary tokens (RID and SID) performs poorly in both diversity and memorization dimensions. While the use of OOV tokens (CID) may improve diversity, its memorization score leaves much to be desired. This suggests the need to explore better explore better forms of OOV tokens to construct IDs, to make them capture user/item correlations while also distinguishing different items.

Our quantitative assessment, based on the metrics we introduced, reveals intriguing insights into the quality of these ID representations. Figure~\ref{fig:score} illustrates that IDs constructed from in-vocabulary tokens (RID and SID) perform poorly in terms of both diversity and memorization. Although the use of OOV tokens (CID) enhances diversity, its memorization score is unsatisfactory. These findings highlight the need to further develop more effective forms of OOV tokens for ID construction, aiming to improve LLMs' ability to capture user/item correlations and distinguish items.

%% file: method.tex
\section{META ID}
\label{Section5}
\begin{figure}[t!]
    \centering
    \subfloat[META ID construction for users and items.\label{fig:method_a}]{
        \includegraphics[width=0.43\linewidth]{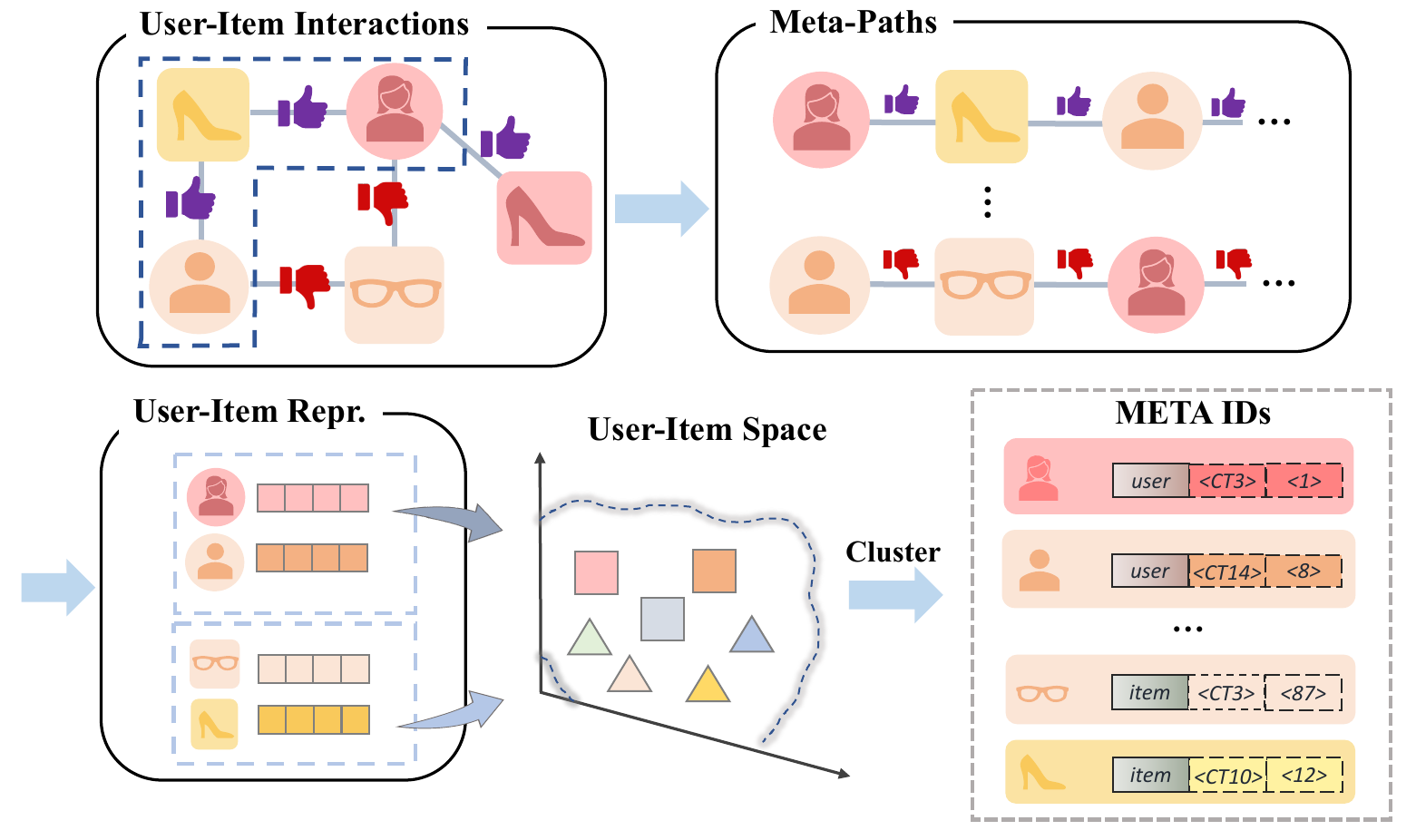}
    }
    \subfloat[Integration of META ID for sequential recommendation.\label{fig:method_b}]{
        \includegraphics[width=0.56\linewidth]{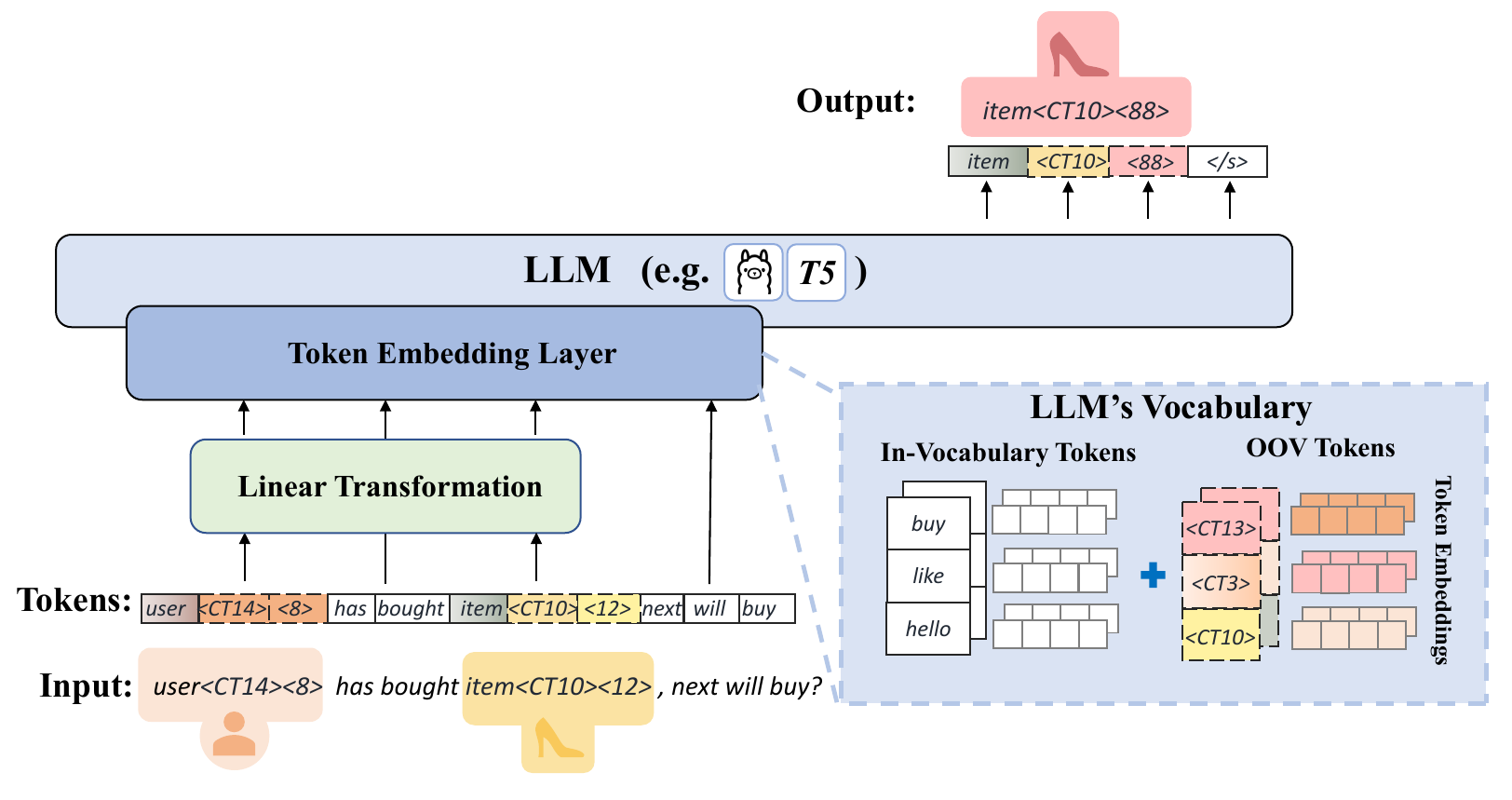}
    }
  \caption{Illustration of the proposed framework: (a) The first step involves sampling meta-paths to train a skip-gram model to learn the representation of users and items. Following this, users and items are represented by META IDs constructed of OOV token combinations through clustering their representations. (b) A large language model (LLM) integrated with META ID is utilized for sequential recommendation. The LLM first encodes the input sequence into tokens and then lookup their embeddings through the token embedding layer. Here, the OOV tokens undergo an extra transformation layer before the token embedding layer of LLMs as a representation augmentation.}

\label{fig:method}
\end{figure}

 We now introduce the META ID framework to enhance LLMs in recommendation tasks. META ID involves creating out-of-vocabulary (OOV) tokens for constructing user and item IDs, which provide a rich, expressive space and encapsulate comprehensive, collaborative information. As illustrated in Figure \ref{fig:method}, our process begins with sampling meta-paths from user-item interaction history. We then apply a skip-gram model to learn user and item representations from these meta-path sequences. This process ensures that users' and items' features are projected into a shared space to capture their interaction relationship better. This is followed by K-Means clustering to group similar users or items, after which unique OOV tokens are assigned to each cluster to construct the META ID.

\subsection{Meta-path-based Embedding}
\label{Section5-1}
We frame user-item interactions within a graph embedding learning paradigm~\cite{yuxiao2017}, constructing an interaction graph composed of user nodes $U$ and item nodes $I$, linked by interaction history with ratings $R$. The core of this embedding learning strategy involves meta-paths~\cite{yuxiao2017} --- a sequence of connections that reflect composite relationships within the graph. Our primary meta-path, denoted as \textit{U-I-U}, materializes when users consistently rate an item with the same score $R_i$, forming a path:
\begin{equation}
\label{meta-path}
\small
    p = \{U_1 \xrightarrow{R_i} I_2 \xrightarrow{R_i} U_3 \xrightarrow{R_i} I_4... \xrightarrow{R_i} U_k \}^{|R|}_{i=1}.
\end{equation}

We employ a skip-gram model to capture the interaction dynamics represented by these meta-paths. This model is trained on sequences generated from meta-path-based random walks~\cite{yuxiao2017}, producing representations $\mathbf{W_U} = [\mathbf{w}_{u_1}^\top, ..., \mathbf{w}_{u_m}^\top ]\in \mathbb{R}^{d\times m}$ and $\mathbf{W_I} = [\mathbf{w}_{i_1}^\top, ..., \mathbf{w}_{i_n}^\top ]\in \mathbb{R}^{d\times n}$, denoted the representations for $m$ users and $n$ items in a $d$-dimension space respectively. Through this process, we acquire deep representations of user and item interactions, which are instrumental in enhancing the accuracy and relevance of our recommendation system.

\subsection{OOV Token Generation}
\label{Section5-2}
To balance diversity and memorization in large-scale recommendation systems, we adopt a cluster-based approach. Representations derived from user-item interactions are organized into a shared embedding space and segmented into $G$ clusters. Each cluster center, $\mathbf{\mu}_g$, is defined by the average of representations within that cluster, effectively capturing the collective characteristics of its members. These centroids then categorize each user and item, providing a refined foundation for constructing granular IDs. The generation process is a two-step procedure: 

{\bf 1). Assign coarse-grained tokens based on centroids}. We first cluster the learned representations $\textbf{W}$ and set the number of clusters as $G$. Each cluster centroid $\mathbf{\mu}_g$ is calculated  as: 
\begin{equation}
\label{eq:9}
\small
    \mathbf{\mu}_g = \left\{\frac{1}{|\mathbb{I}(g_i = g)|}\sum_{(\mathbf{w}_i, g_i) \in (\mathbf{W}, G)}{\big[\mathbf{w}_i \cdot \mathbb{I}(g_i = g)\big]}\right\}_{g \in [1, G]},
\end{equation}
in which we use $i$ to represent a user/item for simplification. As a coarse-grained distinction between users and items, we assign an OOV token ``$\left \langle CT_i \right \rangle$'' to each centroid.

{\bf 2). Assign fine-grained tokens based on distance}. Within each cluster, assign a token based on the distance to the centroid. In detail, we assign fine-grained tokens ``$\left \langle y_i \right \rangle$'' in ascending order according to their distance from the cluster centroid, and use it as a fine-grained token.

The resulting identifiers, META ID, combine a coarse-grained token and a fine-grained one that uniquely identifies each user or item within that cluster. For example, an item might be represented as ``$\left\langle \text{{Item}} \right \rangle \left \langle CT_i \right \rangle \left \langle y_i \right \rangle$'' labeled with three tokens, ``$\left \langle \text{{Item}} \right \rangle$'', ``$\left \langle CT_i \right \rangle$'', ``$\left \langle y_i\right \rangle$'', where ``$\left \langle \text{{Item}} \right \rangle$'' denotes it as an item, ``$\left \langle CT_i \right \rangle$'' is its coarse-grained token and ``$\left \langle y_i \right \rangle$'' is its fine-grained token.

This clustering and labeling process effectively compresses the vocabulary needed for ID representation while preserving the rich information necessary for recommendation tasks. In our implementation, we apply K-Means clustering~\cite{david2007kmeans}, utilizing cosine similarity to measure the affinity of representations to cluster centers. This method simplifies the complexity of representations space management and proves to be robust in our experimental validations of Section~\ref{Section6-4}.

\subsection{Integration of META ID with LLMs}
\label{Section5-3}
The integration of OOV tokens of META ID, denoted as $\mathbf{x_{\text{OOV}}}$, with LLMs involves expanding the vocabularies. This is achieved by extending the token embedding layer's parameters from $\mtheta_{\boldsymbol{E}} \in \mathbb{R}^{N \times d}$ to $\mtheta_{\boldsymbol{E}'} \in \mathbb{R}^{(N+n) \times d}$, where $N$ is the number of in-vocabulary tokens, $n$ is the number of OOV tokens, and $d$ is the dimension of the token embeddings.

A good initialization helps token embedding learning, here we give a representation augmentation approach different from previous works using random initialization~\cite{GengRecommendation22,wenyue2023}. As shown in Figure~\ref{fig:method_b}, the OOV tokens undergo a linear layer $\boldsymbol{F}(\cdot)$ initialized with the category embeddings $\mathbf{\mu}_g$ from Equation~\ref{eq:9}. Finally, the training objective for integrating META ID into LLMs is reformulated as:
\begin{equation}
\small
\mtheta'^* = \operatorname*{arg\,min}_{(\mtheta', \boldsymbol{F})} \mathcal{L}_{\mtheta'} = -\sum_{j=1}^{|\mathbf{y}|} \log P_{\theta'}\big(y_j \mid y_{<j}, \boldsymbol{E}'(\mathbf{x}), \boldsymbol{F}(\mathbf{x_{\text{OOV}}})\big),
\end{equation}
where $\mtheta'$ now includes $\boldsymbol{F}$, aligning with our modified embedding layer $\mtheta_{\boldsymbol{E}'}$ to optimize the model's performance with the OOV tokens $\mathbf{x_{\text{OOV}}}$. This training goal ensures the model can effectively distinguish between diverse users and items, improving its ability to capture user-item relationships. By enhancing memorization and diversity of OOV tokens, the model achieves better performance in recommendation tasks, leading to more accurate and personalized recommendations in our experiments.

%% file: experiments.tex
\begin{table}[t!]
\setlength{\tabcolsep}{2pt}
\small
  \caption{\small Performance comparison of different methods on sequential recommendation task. META ID (T) and META ID (L) refer to the use of T5 and LLaMA2-7b as the backbone. The best and second-best performance methods are denoted in bold and underlined fonts respectively.}
\begin{adjustbox}{width=\linewidth}\
\begin{tabular}{lcccccccccccc}
\toprule
\multirow{2.5}{*}{Methods} & \multicolumn{4}{c}{\textbf{Sports}} & \multicolumn{4}{c}{\textbf{Beauty}} &  \multicolumn{4}{c}{\textbf{Toys}} \\
\cmidrule(lr){2-5}\cmidrule(lr){6-9}\cmidrule(lr){10-13}
 & H@5 & N@5 & H@10 & N@10 & H@5 & N@5 & H@10 & N@10 & H@5 & N@5 & H@10 & N@10 \\
\cmidrule{1-13}
Caser \cite{jiaxi2018}   &  0.0116  &  0.0072  &  0.0194 &  0.0097 & 0.0205 &  0.0131& 0.0347 & 0.0176 &  0.0166 &  0.0107 &  0.0270 &  0.0141  \\
HGN \cite{chen2019}    &  0.0189  &  0.0120  &  0.0313 &  0.0159 & 0.0325 & 0.0206 & 0.0512 & 0.0266 & 0.0321  & 0.0221  &  0.0497 &  0.0277  \\
GRU4Rec \cite{bal2016}  &  0.0129  &  0.0086  &  0.0204 &  0.0110 & 0.0164 & 0.0099 & 0.0283 & 0.0137 &  0.0097 &  0.0059 &  0.0176 &  0.0084  \\
BERT4Rec \cite{fei2019}  &  0.0115  &  0.0075  &  0.0191 &  0.0099 & 0.0203 & 0.0124 & 0.0347 & 0.0170 & 0.0116  &  0.0071 & 0.0203  &  0.0099  \\
FDSA  \cite{yong2023}  &  0.0182  &   0.0122 &  0.0288 &  0.0156 & 0.0267 & 0.0163 & 0.0407 & 0.0208 & 0.0228  & 0.0140  &  0.0381 &  0.0189  \\
SASRec \cite{wang2018}  & 0.0233 & 0.0154 & 0.0350 & 0.0192 &  {0.0387}  &  {0.0249}  & {0.0605}  & {0.0318} & \underline{0.0463}  & \underline{0.0306}  & {0.0675}  &  {0.0374}  \\
S$^3$-Rec \cite{kun2020}  & 0.0251 & 0.0161 & 0.0385 & 0.0204 &  {0.0387}  &  0.0244  & {0.0647}  & 0.0327 & {0.0443}  & {0.0294}  & \underline{{0.0700}}  &  \underline{0.0376}  \\
CL4SRec \cite{xu2022contrastive} & 0.0219	&0.0138&	0.0358&	0.0182	&0.0330	&0.0201&	0.0546&	0.0270	&0.0427	&0.0244&	0.0617	&0.0305\\

\midrule
RID \cite{wenyue2023} &  0.0208  &  0.0122  &  0.0288 &  0.0153 & 0.0213 & 0.0178 & 0.0479 & 0.0277 &  0.0044 & 0.0029  &  0.0062 &  0.0035  \\
SID \cite{GengRecommendation22}  &  0.0223  &  0.0173  &  0.0294 & 0.0196 & 0.0404 & {0.0299} & 0.0573 & {0.0354} &  0.0050 &  0.0031 & 0.0088  &  0.0043  \\
CID \cite{wenyue2023}  &  {0.0269}  &  {0.0196}  &  0.0378 & {0.0231} & 0.0336 & 0.0227 & 0.0507 & 0.0281 &  0.0172 &  0.0109 & 0.0279  &  0.0143  \\
\cmidrule{1-13}
\textbf{META ID (T)}  & \underline{0.0322} &  \underline{0.0223} &  \underline{0.0487} & \underline{0.0277} &  \textbf{{0.0510}}  &  \textbf{0.0351}  & {\textbf{0.0753}} &  \textbf{0.0429} & \textbf{{0.0503}}  &  \textbf{0.0352} &  \textbf{0.0742} &  \textbf{0.0429}  \\

\textbf{META ID (L) }  & \textbf{0.0392} &  \textbf{0.0278} &  \textbf{0.0561} & \textbf{0.0332} &  \underline{{0.0458}}  &  \underline{0.0320}  & {\underline{0.0678}} &  \underline{0.0391} & {{0.0387}}  &  {0.0264} &  {0.0535} &  {0.0312}  \\

\bottomrule
\end{tabular}
\end{adjustbox}
\vspace{-10pt}
\label{tab:seq-rec}
\end{table}

\section{Experiment}

We evaluate META ID on five downstream recommendation tasks: sequential recommendation, direct recommendation, rating prediction, explanation generation, and review related tasks. We analyze the influence of critical components in META ID and assess the ID representations through visualization and our proposed metrics. Details of task descriptions and pre-processing are in Appendix \ref{appendix3}.

\subsection{Evaluation on Sequential Recommendation}
\label{Section6-1}
\noindent{\bf Setups}. We evaluate our META ID framework on three public real-world datasets from the Amazon Product Reviews dataset \cite{ni2019}, focusing specifically on Sports, Beauty, and Toys. The datasets are processed following the methodology in P5~\cite{GengRecommendation22}. 

\noindent{\bf Baselines}. We compare to a variety of established models (which are described briefly in Appendix \ref{appendix3}), spanning from CNN-based to LLM-based frameworks. Caser \cite{jiaxi2018}, HGN \cite{chen2019}, GRU4Rec \cite{bal2016}, BERT4Rec \cite{fei2019}, FDSA \cite{yong2023}, SASRec \cite{wang2018}, S$^3$-Rec \cite{kun2020} and CL4SRec \cite{xu2022contrastive}. Specifically, we provide P5 with its variations, equipped with different ID construction strategies like Sequential ID (SID), Random ID (RID), and Collaborative ID (CID)~\cite{wenyue2023}. 

\noindent{\bf Evaluations}. We apply widely accepted metrics, top-k Hit ratio (H@K) and Normalized discounted cumulative gain (N@K) with K = 5, 10 to evaluate the recommendation performance.

\noindent{\bf Implementation Details}. For constructing META IDs, the clustering groups are limited to $|G|$=100 (200 for Toys). For LLM fine-tuning, we consider both encoder-decoder architecture T5-small \cite{colin2020} and decoder-only architecture LLaMA2-7b \cite{hugo2023llama}. We fully fine-tune the T5 model and employ the LoRA \cite{edward2022lora} to fine-tune LLaMA2-7b. Vocabulary sizes of these models are shown in Table~\ref{tab:token_size}. For LLM inferencing, we use beam search to generate potential items evaluated under the all-item setting.

\noindent{\bf Results}. Table \ref{tab:seq-rec} presents our findings for sequential recommendation task~\footnote{We show the standard error of the metrics for META ID in Table~\ref{tab:br}.}. Our observations are as follows: 1) META ID demonstrates superior performance on all three datasets, underscoring its robustness. 2) IDs constructed of in-vocabulary tokens, RID and SID, underperform on Toys, suggesting limitations in their recommendation efficacy for LLMs. 3) CID shows marked improvements over RID and SID on Toys dataset, highlighting the benefits of incorporating OOV tokens with collaborative information. 4). While the LLaMA2-7b backbone is better in the Sports dataset, its performance in the Beauty and Toys dataset is not as good as T5, which could be linked to the distinct fine-tuning methodologies applied to these models.

\subsection{Evaluation on Various Recommendation Tasks}

\label{Section6-2}
\noindent{\bf Setups}. To validate META ID's adaptability, we extend our evaluation to include direct recommendation, rating prediction, explanation generation, and review tasks, akin to P5~\cite{GengRecommendation22}. For direct recommendation, the model is asked to recommend item for users directly without providing user's interaction history. For rating prediction, the model predicts a numerical rating between 1 and 5 based on user-item data. For explanation tasks, it generates textual justifications for a user's preference towards an item, while in review tasks, it summarizes lengthy reviews into concise titles. 

\noindent{\bf Baselines}. We compare to three different ID construction strategies: RID, SID, and CID. 

\noindent{\bf Evaluations}. For direct recommendation, we apply the same metrics as in Section~\ref{Section6-1}. For rating prediction, we use MSE metric. For explanation and review tasks, we employ BLEU-1/4 metrics.\looseness=-1

\noindent{\bf Implementation Details}. For inferencing, we apply greedy decoding for rating, explanation, and review tasks, and beam search under the all-item setting for direct recommendation task.

\noindent{\bf Results}. 
For direct recommendation (Table \ref{tab:direct}), META ID exceeds other methods across datasets in all-item setting. This suggests that META ID effectively model the direct relationship between users and items. The results for the other three tasks in Table \ref{tab:rating}, show that META ID significantly improves the BLEU scores Sports and Beauty compared to other methods. This result suggests that META ID can improve performance in text relevance tasks, including the interpretation of recommendations.

\begin{table*}[t!]

  \caption{\small Performance comparison of different methods on direct recommendation task. }
  \label{tab:direct-rec}
  \setlength{\tabcolsep}{0.8mm}
  \begin{adjustbox}{width=\linewidth}
\begin{tabular}{lcccccccccccc}
\toprule
\multirow{2.5}{*}{Methods} & \multicolumn{4}{c}{\textbf{Sports}} & \multicolumn{4}{c}{\textbf{Beauty}} &  \multicolumn{4}{c}{\textbf{Toys}} \\
\cmidrule(lr){2-5}\cmidrule(lr){6-9}\cmidrule(lr){10-13}
 & H@5 & N@5 & H@10 & N@10 & H@5 & N@5 & H@10 & N@10 & H@5 & N@5 & H@10 & N@10 \\
\cmidrule{1-13}
RID~\cite{wenyue2023} &  0.0030 & 0.0023  & 0.0042  & 0.0027 & 0.0203 & 0.0155 &0.0276 & 0.0178 & 0.0046  & 0.0030  & 0.0063  &  0.0035  \\
SID~\cite{GengRecommendation22}  &  0.0211  &  0.0169  &  0.0267 & 0.0187 &  \underline{0.0296} & \underline{0.0226} & \underline{0.0405} & \underline{0.0261} &  0.0025 & 0.0014  &  0.0041 &  0.0019  \\
CID~\cite{wenyue2023}    &  \underline{0.0250}  &  \underline{0.0189} &  \underline{0.0342} & \underline{0.0219} &  0.0216 & 0.0147 & 0.0340 & 0.0187 &  \underline{0.0076} & \underline{0.0049}  &  \underline{0.0014} &  \underline{0.0070}  \\
\textbf{META ID}  &  \textbf{0.0357}  &  \textbf{0.0256}  &  \textbf{0.0520} & \textbf{0.0308}  &\textbf{ 0.0480} &\textbf{ 0.0336} & \textbf{0.0689} & \textbf{ 0.0403} &  \textbf{0.0564} & \textbf{0.0391}  & \textbf{0.0803}  &  \textbf{0.0468}  \\
\bottomrule
\end{tabular}
\end{adjustbox}
\vspace{-15pt}
\label{tab:direct}
\end{table*}

\begin{table}[t]
  \caption{\small Performance comparison of different methods on rating prediction, explanation generation, and review tasks. Due to space constraints, see Table~\ref{tab:rating-app} for full results.}
  \label{tab:rating}
  \setlength{\tabcolsep}{1.2mm}
  \small
  \begin{adjustbox}{width=\linewidth}
  \begin{tabular}{lccccccccc}
    \toprule
    \multirow{2}{*}{Task Type} & \multirow{2}{*}{Metric} & \multicolumn{4}{c}{\textbf{Sports}} & \multicolumn{4}{c}{\textbf{Beauty}}  \\
    \cmidrule(lr){3-6} \cmidrule(lr){7-10}
     & & RID & SID & CID & \textbf{META ID} & RID & SID & CID & \textbf{META ID} \\
    \midrule
    \multirow{1}{*}{Rating} & RMSE 
        & \underline{1.0382} & 1.0486 & 1.0383 & \textbf{1.0327} 
        & 1.2829 & 1.3098 & \underline{1.2819} & \textbf{1.2818}  \\
    \midrule
    \multirow{2}{*}{Explanation} & BLEU-1 
        & 16.2567 & 16.5825 & {16.6121} & \textbf{16.9005} 
        & 18.2299 & 18.3981 & \underline{19.3499} & \textbf{19.5106}\\
                                 & BLEU-4 
        & 2.1782 & 2.1944 & \underline{2.2332} & \textbf{2.3481} 
        & 2.9027 & 2.8071 & \textbf{3.0626} & \underline{3.0592}\\
    \midrule
    \multirow{2}{*}{Review} & BLEU-1 
        & 7.6140 & \underline{7.7948} & 7.6586 & \textbf{7.8819} 
        & 6.2282 & 6.5055 & \underline{6.5854} & \textbf{7.0500}\\
                            & BLEU-4 
        & 2.3228 & 1.2406 & \underline{2.4109} & \textbf{2.6546} 
        & \underline{1.9891} & 1.2406 & 1.9718 & \textbf{2.7485}\\
    \bottomrule
  \end{tabular}
  \end{adjustbox}
  \vspace{-10pt}
\end{table}

\subsection{Evaluation of ID Representation}

\label{Section6-3}
\noindent{\bf Visualization}. The amount of numeric tokens available in LLMs is relatively limited, which complicates the establishment of unique one-to-one ID relationships, and two unrelated items might share the tokens as ID. For an intuitive explanation, we visualized the cosine similarity matrix between items using heatmaps in Figure \ref{fig:teaser}, where we random sample 50 items from the Toys dataset and take their adjusted cosine similarity from Equation~\ref{eq:adjusted_sim} as ground truth compared with RID, SID and META ID. RID and SID result in a large number of similar items due to semantic conflicts, while META ID shows distinguished similarity closer to ground truth. This suggests that using META ID allows LLMs to better capture relationships between users and items. 

\noindent{\bf Quantitative Analysis}. We quantitatively assess the quality of these ID representations by the proposed two metrics (Section \ref{Section3-3}): memorization score (MS) and diversity score (DS). Our results, shown in Figure \ref{fig:score}, indicate that constructing IDs of in-vocabulary tokens (RID and SID) perform poorly in the diversity dimension. For intuitive interpretation, we further employ t-SNE visualization to map ID representations and observe a tendency for these tokens to cluster narrowly in Figure~\ref{fig:score}. META ID shows robust memorization and diversity across three datasets, reflecting its ability to\begin{wrapfigure}{r}{0.5\textwidth}
\vspace{-5pt}
    \centering
    \includegraphics[width=7cm]{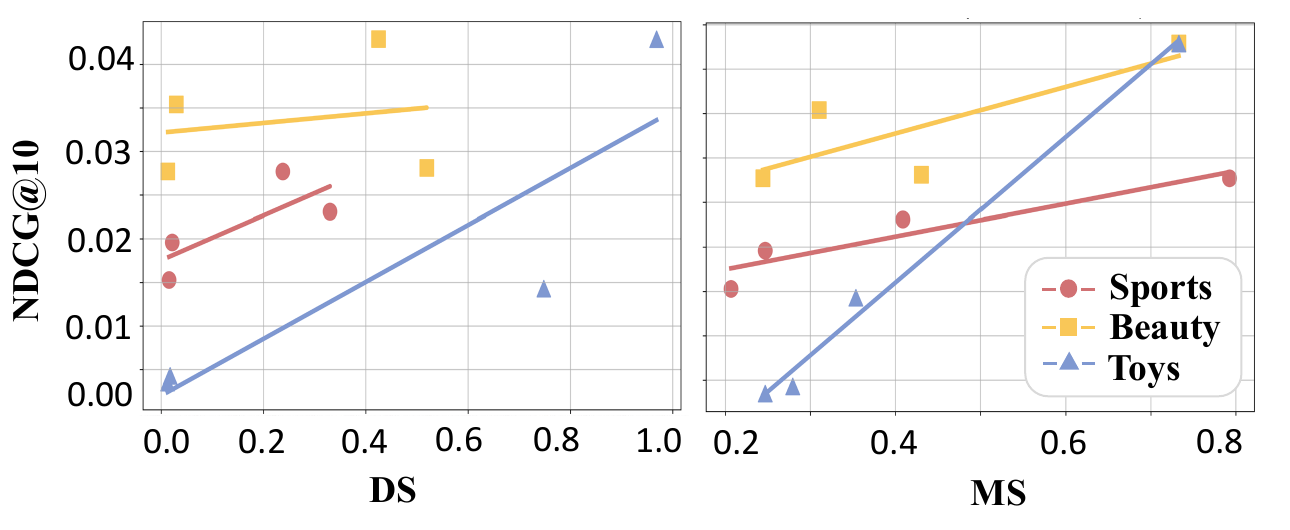}
  \caption{ \small Correlation between DS and MS with NDCG@10 on sequential recommendation task.}
\vspace{-15pt}
\label{fig:score_b}
\end{wrapfigure} capture correlations between users and items from historical data while ensuring items remain distinguishable.

\noindent{\bf Metrics Analysis}. Furthermore, we conduct a correlation analysis to explore the relationship between MS/DS and sequential recommendation performance. We sum the MS and DS of different ID strategies and plot their performance on the sequential recommendation task. As shown in Figure \ref{fig:score_b}, the sum of the MS and DS positively correlate with NDCG@10, suggesting that memorization and diversity of IDs are two essential properties in recommendation tasks.

\subsection{Ablation Studies}
\label{Section6-4}
We analyze the properties of META ID following the evaluation in Section \ref{Section6-1}, including the impact of\begin{wrapfigure}{r}{0.4\textwidth}
    \centering
    \vspace{-5pt}
    \subfloat[Different Size of OOV tokens.\label{fig:token_a}]{
        \includegraphics[width=5.5cm]{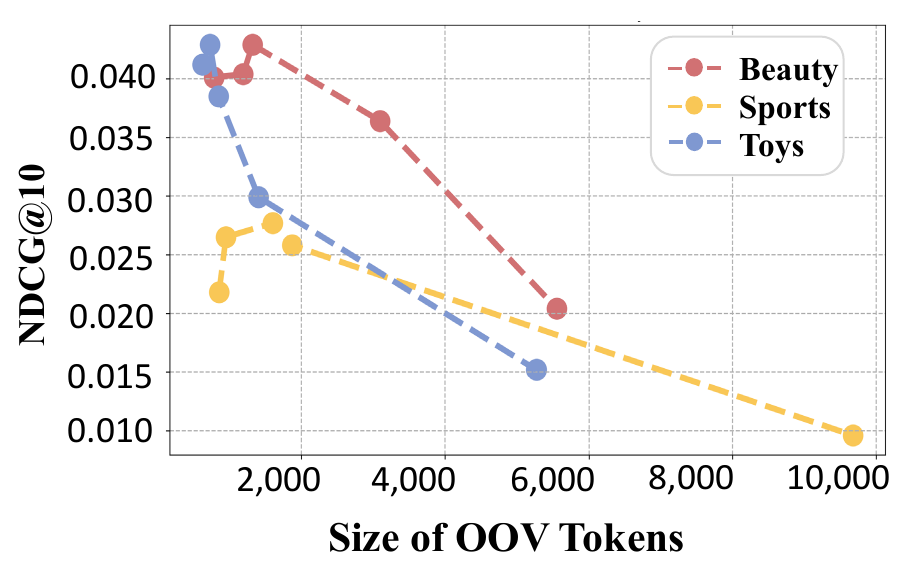}
    }
    \qquad
    \subfloat[Different Indexing Range.\label{fig:token_b}]{
        \includegraphics[width=5.3cm]{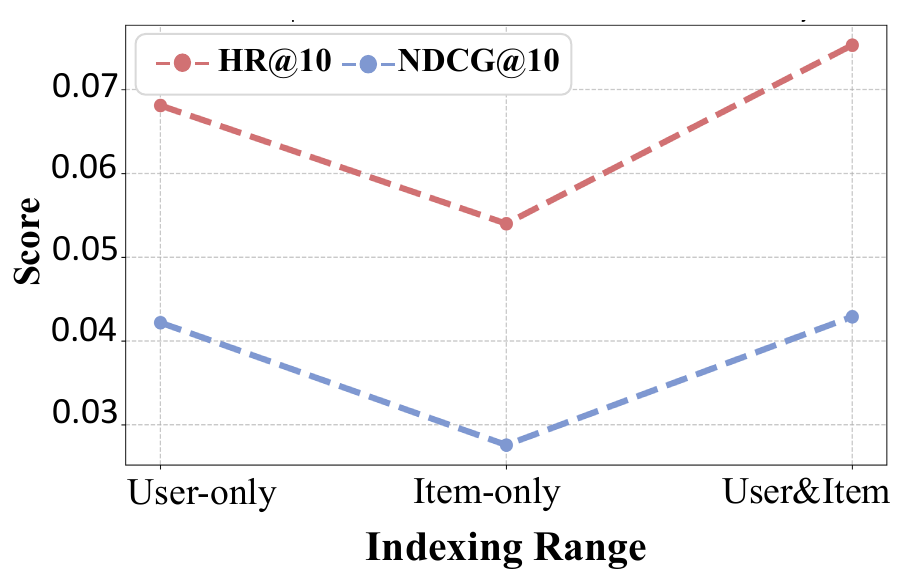}
    }
  \caption{ \small Performance comparison of (a) different OOV token size within META ID on three datasets; (b) whether indexing both items and users on Beauty dataset.}
\vspace{-25pt}
\end{wrapfigure} grouping methods, the size of OOV tokens, and different indexing ranges on the performance of META ID.

\begin{table*}[t!]
\vspace{-5pt}
  \caption{\small Performance of different grouping methods for META ID on sequential recommendation task.}
  \label{tab:group}
  \setlength{\tabcolsep}{0.8mm}
  \begin{adjustbox}{width=\linewidth}
  \begin{tabular}{ccccccccccccc}
    \toprule
    \multirow{2.5}{*}{Methods} & \multicolumn{4}{c}{\textbf{Sports}} & \multicolumn{4}{c}{\textbf{Beauty}} &  \multicolumn{4}{c}{\textbf{Toys}} \\
    \cmidrule(lr){2-5}\cmidrule(lr){6-9}\cmidrule(lr){10-13}
   & H@5 & N@5 & H@10 & N@10 & H@5 & N@5 & H@10 & N@10 & H@5 & N@5 & H@10 & N@10 \\
    \cmidrule{1-13}

    DBSCAN \cite{martin1996a} & 0.0078 & 0.0043 & 0.0145 & 0.0065 & 0.0257 & 0.0168 & 0.0428 & 0.0223 & 0.0180  & 0.0109 & 0.0304 & 0.0149 \\
    Spectral \cite{kluger2003spectral} & 0.0199 & 0.0124 & 0.0336 & 0.0167 & 0.0360  & 0.0236 & 0.0588 & 0.0310 & 0.0295 & 0.0184 & 0.0514 & 0.0254 \\
    RQ-VAE \cite{neil2022sound} & 0.0122 & 0.0077 & 0.0171 & 0.0093 & 0.0368 & 0.0254 & 0.0536 & 0.0309 & \textbf{0.0511} & 0.0335 & 0.0667 & 0.0395 \\
    K-Means \cite{david2007kmeans} & \textbf{0.0322} & \textbf{0.0223} & \textbf{0.0487} & \textbf{0.0277} & \textbf{0.0510} & \textbf{0.0351} &\textbf{ 0.0753} & \textbf{0.0429} & 0.0503 & \textbf{0.0352} & \textbf{0.0742} & \textbf{0.0429} \\
    \bottomrule
  \end{tabular}
  \end{adjustbox}
\vspace{-15pt}
\end{table*}

\noindent{\bf Token Grouping}.
We study the importance of different grouping methods for OOV token generation in our framework. In Table \ref{tab:group}, we compare the performance of DBSCAN \cite{martin1996a} and Spectral Clustering \cite{kluger2003spectral} against K-Means clustering \cite{david2007kmeans}. Since related work \cite{shashank2023recommender} has not yet opened source code, here we implement it ourselves by generating OOV tokens with RQ-VAE \cite{neil2022sound} using meta-path-based embeddings. Our results show that simply applying K-Means outperforms other grouping methods in most cases. 

\noindent{\bf OOV Token Size}. Since varying cluster sizes $G$ result in different numbers of OOV tokens, we also investigate the impact of different cluster sizes for META ID in Figure~\ref{fig:token_a}. We find that the granularity of token clusters plays a crucial role in recommendation performance. An excessive token scale can introduce noise, reducing the performance. Therefore, finding an optimal token size is vital to ensure that META ID effectively adapts to various datasets' nuances.

\noindent{\bf User or Item Indexing}.
Previous ID strategies for LLMs only consider indexing for items, which come from the convention that users are typically represented by a sequence of interacted items in sequential recommendation~\cite{GengRecommendation22,wenyue2023,shashank2023recommender}. While META ID models users and items, as shown in Table~\ref{fig:token_b}, reveals that the combined user-item indexing (User\&Item) outperforms either user-only or item-only indexing. This result shows the importance of incorporating user preferences and item attributes for LLMs to enhance the accuracy of the recommendations.

%% file: conclusion.tex
\section{Conclusion}
\label{Section7}
This study introduces META ID, a method enhancing Large Language Models (LLMs) for recommender systems using OOV tokens. Moving beyond constructing IDs with in-vocabulary tokens, META ID incorporates user-item interaction information to align LLMs more effectively with recommendation tasks. We learn representations from user-item interactions utilizing meta-paths sampling. By clustering these representations we generate OOV tokens to construct META ID. This approach guarantees tokens capturing correlations between users and items from historical data while ensuring distinctiveness among item. Our experiments across various real-world datasets demonstrate META ID's robust performance in diverse recommendation tasks, including sequential and direct recommendation, as well as complex tasks requiring detailed textual responses.
Essentially, META ID effectively combines the capabilities of LLMs with the nuanced requirements of recommendation scenarios, such as planning highly personalized content for users as virtual shopping assistants.

%% file: appendix.tex
\newpage
\appendix
\begin{center}
    {\large \textbf{Appendix}}
\end{center}
We provide details omitted in the main paper. 
\begin{itemize}[itemsep=1pt,topsep=0pt]
    \item \autoref{appendix1}: Workflow of META ID, encompassing the construction of OOV tokens.
    \item \autoref{appendix2}: Details of memorization score (MS) and diversity score (DS).
    \item \autoref{appendix3}: Experimental setups and implementation details of META ID.
    \item \autoref{appendix4}: Additional experimental result analysis.
    \item \autoref{appendix5}: Discussions and limitations of META ID.
\end{itemize}

\section{Details of META ID}
\label{appendix1}
In the Section~\ref{Section5} of the main text, we elucidate the comprehensive workflow for generating META ID. This process encompasses three main steps, including (1) the extraction of meta-path-based embedding, (2) the generation of OOV tokens, and (3) the incorporation of META ID with LLMs, thereby handling with various downstream recommendation tasks. 

\subsection{How to extract the meta-path-based embedding}
This section supplements the details of subsection~\ref{Section5-1}, \ie, the users / items representations extracted from a skip-gram model, including the sampling process of meta-paths as training data. 

In META ID, we enable a skip-gram model to learn effective users / items representations from the sampled meta-paths, which is learning user representations $\mathbf{W}_U$ and item representations $\mathbf{W}_I$. The objective of the skip-gram model learning paradigm is to map the users and items in the meta-paths seqeuences into a lower-dimensional space as in~\cite{yuxiao2017}

\noindent{\bf Meta-paths sampling}. Firstly, we constructs a node sequence based on random walks of meta-paths. A meta-path $p = P_1 \xrightarrow{R_1} P_2 \xrightarrow{R_2} ... \xrightarrow{R_{k-1}} P_k $ is a path that is defined on a graph, where $R_i$ signifies a composite relation between different node $P$. We define user-item-user (\textit{U-I-U}) as our meta-path, where paths only exist if users has given the same ratings $R_i$ to one item. We sample 32 rounds starting from each user and item with the sampled length $k=64$.

\noindent{\bf Skip-gram model training}. In the second step, through sampling meta-paths based on random walks as training corpus, we train a skip-gram model thus learn the vector representations $(\mathbf{W}_U, \mathbf{W}_I)$ for all users and items. The objective of the skip-gram model is to maximize the conditional probability $P(n_i|v)$ for the node $v \in V$ of its neighboring node $n_i \in N_v$:
\begin{equation}
\small
    \arg\max_{\theta} \sum_{v\in V} \sum_{t\in {U, I}} \sum_{n_i\in N_v} \log P(n_i|v)
\end{equation}
and the $P(n_i|v)$ is calculated as:
\begin{equation}
\small
    P(n_i|v)=\frac{\exp(w_{n_i}^T w_v)}{\sum_{j\in V}\exp(w_j^T w_v)}
\end{equation}
where $w$ means the representation of one user or item. In experiments, we set number of negative sampling to 5, and train the skip-gram model for 10 epoch with learning rate of 0.001.

\subsection{How to generate OOV tokens}
This section complements subsection \ref{Section5-2}, where we generate OOV tokens from users and items representations for constructing META ID. Essentially, we need to build a hierarchical classification system for IDs in order to express a wider range of items and users with as few OOV tokens as possible, so that similar items and users are under the same hierarchical branch.

This hierarchical construction mechanism is very reminiscent of clustering methods, as we apply in META ID, in which we use out-of-class indexes and in-class indexes as two levels of IDs. Though more sophisticated clustering method s for multi-levels structure can be applied, in experiments, we use the simple K-Means clustering, which is more suitable for large-scale data volume due to its simple and easy to optimize nature. We also demonstrate the effectiveness of this approach in Table~\ref{tab:group}. 

In experiments, we cluster user and item representations together. We then create the between-cluster tokens $\left \langle CT_i \right \rangle$ for cluster $i$, and sort the in-cluster users / items based on the cosine distance to the cluster centroids to get in-cluster tokens $\left \langle y_i \right \rangle$. Finally, we add a prefix token $\left \langle \text{\textit{Item}}  \right \rangle$ or $\left \langle \text{\textit{User}} \right \rangle$ to denote its type. For example, an item might be represented as "$\left \langle \text{{Item}} \right \rangle \left \langle CT_i \right \rangle \left \langle j \right \rangle$" labeled with three tokens ("$\left \langle \text{{Item}} \right \rangle$", "$\left \langle CT_i \right \rangle$", "$\left \langle j\right \rangle$"), where "$\left \langle \text{{Item}} \right \rangle$" denotes it as an item, "$\left \langle CT_i \right \rangle$" is its cluster token, and "$\left \langle y_i \right \rangle$" is its in-cluster token.

It is worth noting that a naive approach to generating user and item IDs is to assign an independent OOV token as ID (IID) that needs to be learned for each item and user. However, this is not applicable to modern recommender systems with a large number of items and users, as the training time may be too long if a large number of new tokens need to be created. We also show it in Table~\ref{tab:IID}, where we use the meta-path-based embeddings with a linear projection layer for initialization, which shows that it is ineffective compared to META ID.

\begin{table}[t!]
\centering
\small

\caption{Examples of prompts for recommendation tasks (Part 1).}
\begin{tabular}{lp{0.3\linewidth}p{0.25\linewidth}p{0.25\linewidth}}
\toprule
& \textbf{Sequential Recommendation} & \textbf{Direct Recommendation} & \textbf{Rating Prediction} \\
\midrule
Task Input: & Considering \textbf{user\_2024} has interacted with items \textbf{item\_1}, \textbf{item\_2}. What is the next recommendation for the user? & What should we recommend for \textbf{user\_2024}? & Which star rating will \textbf{user\_2024} give to item \textbf{item\_2}? (1 being the lowest and 5 being the highest). \\
\midrule
Task Output: & \textbf{item\_2024} & \textbf{item\_2024} & 5 \\
\bottomrule
\end{tabular}
\label{tab:prompt}
\end{table}
\begin{table}[t!]
\vspace{-10pt}
\centering
\small
\caption{Examples of prompts for recommendation tasks (Part 2).}
\begin{tabular}{lp{0.35\linewidth}p{0.35\linewidth}}
\toprule
& \textbf{Explanation} & \textbf{Review} \\
\midrule
Task Input: & According to the feature word quality, generate a 5-star explanation for \textbf{user\_2} about \textbf{item\_2}. & Write a short sentence to summarize the following product review from \textbf{user\_2}: Absolutely great product. I bought this for ... \\
\midrule
Task Output: & Absolutely great product! & Perfect! \\
\bottomrule
\end{tabular}
\label{tab:prompt-part2}
\end{table}

\begin{table*}[t!]
\vspace{-5pt}
  \caption{ Comparison of independent ID (IID) with META ID on sequential recommendation task.}
  \label{tab:IID}
  \setlength{\tabcolsep}{0.8mm}
  \begin{adjustbox}{width=\linewidth}
\begin{tabular}{lcccccccccccc}
\toprule
\multirow{2.5}{*}{Methods} & \multicolumn{4}{c}{\textbf{Sports}} & \multicolumn{4}{c}{\textbf{Beauty}} &  \multicolumn{4}{c}{\textbf{Toys}} \\
\cmidrule(lr){2-5}\cmidrule(lr){6-9}\cmidrule(lr){10-13}
 & HR@5  & NDCG@5 & HR@10  & NDCG@10 & HR@5  & NDCG@5 & HR@10  & NDCG@10 & HR@5  & NDCG@5 & HR@10  & NDCG@10  \\
\cmidrule{1-13}

IID    &  {0.0114}  &  {0.0073} &  {0.0208} & {0.0103} &  0.0302 & 0.0194 & 0.0494 & 0.0256 &  {0.0146} & {0.0091}  &  {0.0217} & {0.0114}  \\
\textbf{META ID}  &  \textbf{0.0357}  &  \textbf{0.0256}  &  \textbf{0.0520} & \textbf{0.0308}  &\textbf{ 0.0480} &\textbf{ 0.0336} & \textbf{0.0689} & \textbf{ 0.0403} &  \textbf{0.0564} & \textbf{0.0391}  & \textbf{0.0803}  &  \textbf{0.0468}  \\

\bottomrule
\end{tabular}
\end{adjustbox}
\end{table*}

\subsection{How to incorporate META ID with LLMs}
As mentioned in subsection~\ref{Section3-1}, we convert every recommendation tasks into question\&answering template, in which we describe the recommendation tasks in natural language form, and replace user and item IDs with different dataset like a cloze test. The full templates for every tasks in this format is from~\cite{GengRecommendation22}, where we give some examples in Table~\ref{tab:prompt} and Table~\ref{tab:prompt-part2}.

Take rating prediction task as example. We might ask the LLM, "Which star rating will user $\left \langle \text{User} \right \rangle\left \langle CT_1 \right \rangle\left \langle 18 \right \rangle$  give item $\left \langle \text{Item} \right \rangle\left \langle CT_8 \right\rangle\left\langle 24 \right \rangle$ ?", and expect the LLM to answer "5".

We construct the fine-tuning and testing dataset for LLM in this unified way. Then LLM is able to acquire the generalized knowledge across different tasks, and even carve out user and item characteristics through those tokens constructing IDs to handle different recommendation tasks.

\begin{figure}[t!]
\centering
  \includegraphics[width=\linewidth,height=5cm]{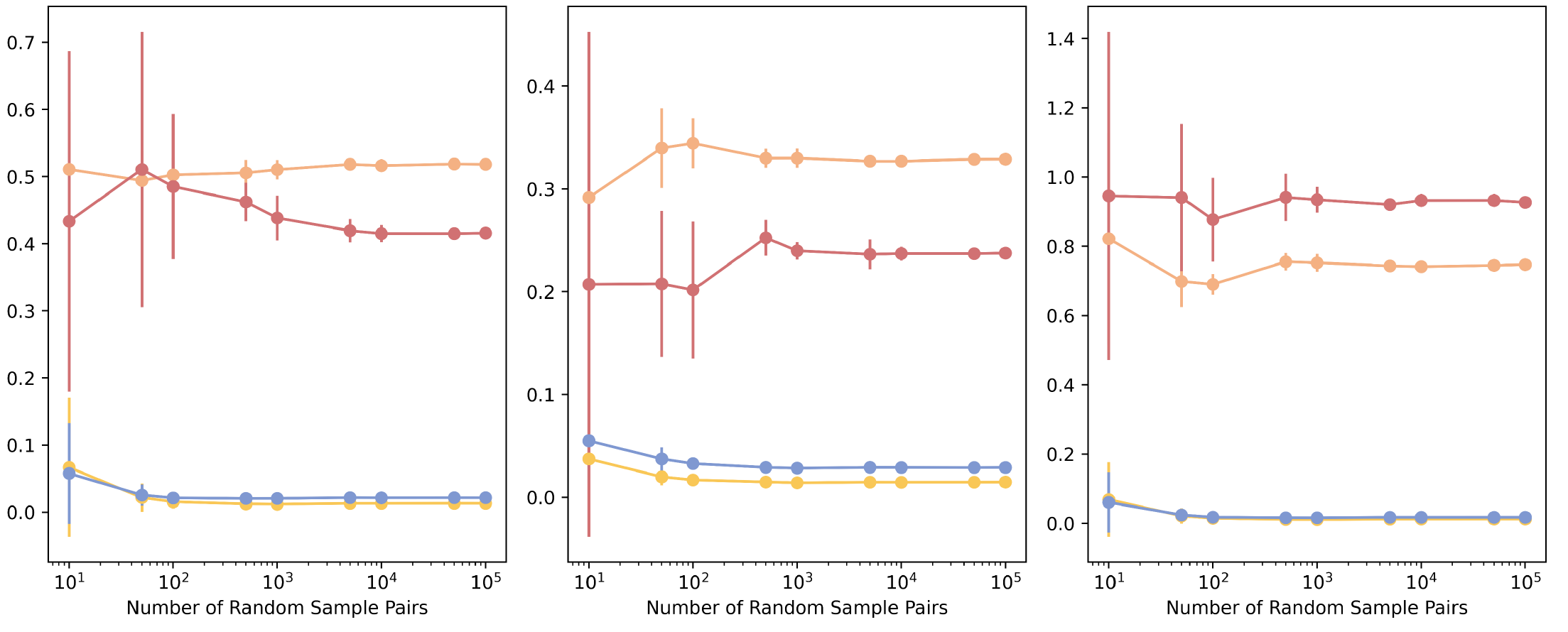}
  \caption{Convergence analysis of DS metric for Beauty and Sports datasets, with 95\% confidence interval error bars indicating variability across 5 trials.}
  % \Description{}
  \label{fig:converge}
\end{figure}

\section{Details of memorization score (MS) and diversity score (DS)}
\label{appendix2}
This section complements subsection~\ref{Section3-3}. The token embedding layer in LLMs transforms each input token into token embedding vectors. And we use ID representation to indicate that these token embedding vectors corresponding to an ID, \ie, the representation of an item or user in LLMs.

\noindent{\bf The convergence of DS}. DS is a metric designed to quantify the diversity of ID representations in LLMs. Given the high computational demand of calculating KL divergence for all embedding pairs in large datasets, DS employs a random sampling approach, thus we present a convergence analysis for DS in Figure \ref{fig:converge}. The stability of DS is evident across both datasets, demonstrating a trend towards convergence as the number of sampled pairs grows. This sampling strategy reducesthe computational complexity from $O(|I|^2 \cdot D)$ to $O(N \cdot D)$, where the $|I|$ means the size of items, $D$ is the dimension of ID representation, which is equal to the LLM’s token embedding dimension.

\noindent{\bf The approximate value of MS}. The adjusted cosine similarity for items is given by:
\begin{equation}
\text{sim}(i, j) = \frac{\sum_{u \in U} (R_{u,i} - \bar{R}_u) \cdot (R_{u,j} - \bar{R}_u)}{ {\sqrt{\sum_{u \in U}{(R_{u,i} - \bar{R}_u)^2} \cdot {\sum_{u \in U}{(R_{u,j} - \bar{R}_u)^2}}}}}
% {\sqrt{\sum_{u \in U}{(R_{u,i} - \bar{R}_u)^2}}}
% {\sqrt{\sum_{u \in U}{(R_{u,j} - \bar{R}_u)^2}}}
\end{equation}
where $R_{u,i}$ and $R_{u,j}$ denote user $u$'s ratings for items $i$ and $j$, respectively, while $\bar{R}_u$ is user $u$'s average rating. To enhance computational efficiency, especially for large-scale datasets, we precalculate the rating deviation sums and squared sums for each item and user:
\begin{equation}
\text{sim}'(i, j) = \frac{\text{Dev}(i) \cdot \text{Dev}(j)}{\sqrt{\text{DevS}(i)} \cdot \sqrt{\text{DevS}(j)}}
\end{equation}
where the rating deviation sums and squared sums for each item is:
\begin{equation}
\text{Dev}(i) = \sum_{u \in U_i} (R_{u,i} - \bar{R}_u),\   \text{DevS}(i) = \sum_{u \in U_i} (R_{u,i} - \bar{R}_u)^2
\end{equation}

This approach reduces complexity from $O(|U|\cdot |I|^2)$ to $O(|I|^2)$, a significant improvement for large-scale datasets.

\section{Experimental Setups and Implementation Details}
\label{appendix3}

\begin{table}[h]
\centering
\caption{Basic statistics of the experimental datasets.}
\begin{tabular}{lrrrr}
\toprule
Dataset &  {\textbf{Sports}} &  {\textbf{Beauty}} & {\textbf{Toys}}\\
\cmidrule{1-4}
\#Users    &  35,598   &  22,363   &  19,412 \\
\#Items    &  18,357   &  12,101  & 11,924\\
\#Reviews  & 296,337   &  198,502   & 167,597 \\
\#Sparsity (\%) &  0.0453  & 0.0734  &  0.0724\\
\bottomrule
\end{tabular}
\label{tab:stats}
\end{table}

\subsection{Datasets Descriptions and Preprocessing}
We conduct extensive experiments over three real-world datasets. The Amazon datasets are collected from \textit{Amazon platform}\footnote{https://nijianmo.github.io/amazon} with user ratings and reviews on 29 categories of products. In this paper, we adopt three of them to evaluate our method, Sports\&Outdoors, Beauty, Toys\&Games. We follow~\cite{GengRecommendation22} and use transaction records between January 1, 2019 to December 31, 2019. Detailed dataset statistics are available in Table \ref{tab:stats}. 

We divide tasks into ratings, explanations, and reviews, adhering to the data-splitting approaches of similar studies~\cite{GengRecommendation22, Lei2020, lei2021}. For both sequential and direct recommendation tasks, we adopt the methodology of \cite{kun2020, GengRecommendation22, ruiyang2020}, using the final item in a user's interaction sequence for testing while carefully structuring the training data to avoid leakage. For rating, explanation, and review task families, we randomly split each dataset into training (80\%), validation (10\%) and testing (10\%) sets, and ensure that there is at least one instance included in the training set for each user and item.

\subsection{Baselines}
Our approach is compared to a variety of established models, spanning from CNN-based to LLM-based frameworks. Caser \cite{jiaxi2018} applies CNNs to capture high-order Markov Chains in sequential recommendation. HGN \cite{chen2019} utilizes hierarchical gating networks for modeling long and short-term user interests. GRU4Rec \cite{bal2016} employs GRUs for session-based recommendation, representing items with embedding vectors. BERT4Rec \cite{fei2019}, S$^3$-rec \cite{kun2020} and SASRec \cite{wang2018} employ self-attention mechanisms for sequential recommendation, focusing on bidirectional understanding and multi-head attention, respectively. FDSA \cite{yong2023} adopts feature-level self-attention for feature transitions. CL4SRec \cite{xu2022contrastive} introduces contrastive learning with data augmentation in sequential recommendation. P5 \cite{GengRecommendation22} learns different tasks with the same language modeling objective during pretraining, serving as the foundation model for various downstream recommendation tasks. P5 \cite{GengRecommendation22} is a recent method that uses a pretrained Large Language Model (LLM) to unify different recommendation tasks in a single model. Since there is no open source code for the recent work~\cite{shashank2023recommender} yet, we implemented the key ID construction of it ourselves in subsection~\ref{Section6-4}.

In particular, we provide P5 with its variations~\cite{wenyue2023}, equipped with different ID constructs like Sequential IDs (SID), Random IDs (RID), and Collaborative IDs (CID) as a benchmark for exploring the impact of different ID strategies. RID Assigns each item with a random number as the item ID. The number is further tokenized into a sequence of sub-tokens, as did in P5. For example, an item is randomly assigned the number "2024", and "2024" is represented as a sequence of tokens "20""24". SID is a straightforward method to leverage collaborative information for item indexing, where items interacted consecutively by a user are assigned consecutive numerical indices, reflecting their co-occurrence. CID approach employs spectral clustering based on spectral matrix factorization to generate item IDs. This method is based on the premise that items with more frequent co-occurrence are more similar and should share more overlapping tokens in ID construction. The results for all baselines except P5 with its variations are reproduced through open source code~\cite{kun2020}.

\subsection{Implementation Details}
As mentioned in Appendix~\ref{appendix1}, we generate the META ID for users and items for each dataset, generalized to all experiments below. For constructing META IDs, we sampling rating-based meta-paths in each dataset, where adjacent users assigned equal ratings to an item. We set the sampling path length to 64, and use a skip-gram model for training, with a window size of 5 and learning rate set at $1e^{-3}$. The embedding clusters groups are limited to $|G|=100$ (200 for Toys) to manage vocabulary size effectively. The OOV tokens size is shown in Table~\ref{tab:token_size}.

\begin{table*}[t!]
\small
\centering
  \caption{Table of the size of OOV tokens used by IDs.}
  \label{tab:token_size}
\begin{tabular}{lccc}
\toprule
OOV Tokens Size & {\textbf{Sports}} & {\textbf{Beauty}} &  {\textbf{Toys}} \\
\cmidrule{1-4}
RID & 0 &0 & 0\\
SID & 0 &0 & 0\\
CID    & 448 &437 &487 \\
IID    & 18357 & 12101 & 11924\\
\textbf{META ID}   & 1600 &1319 & 727\\
\bottomrule
\end{tabular}
\end{table*}

\subsubsection{Evaluation on Sequential and Direct Recommendation Tasks}
We first evaluate META ID on sequential recommendation tasks and direct recommendation tasks following~\cite{wenyue2023}. Our implementation first utilizes T5~\cite{colin2020} as the backbone with parameters around 60.75 million. As mentioned in subsection~\ref{Section5-3}, we add a linear layer where the OOV tokens undergo an extra linear transformation before the token embedding layer for a better initialization with $\alpha=0.1$. We also consider decoder-only architecture LLaMA2-7b~\cite{hugo2023llama} with 7B parameters. For tokenization, we use the default SentencePiece tokenizer with extended OOV tokens for parsing sub-word units. We use the same sequential recommendation and direct recommendation prompts in P5~\cite{GengRecommendation22} to convert sequential information into texts.

\noindent{\bf For LLM fine-tuning}, we pre-train T5 for 10 epochs using AdamW optimizer on two NVIDIA RTX 3090 GPUs with a batch size of 64, a peak learning rate of $1e^{-3}$. We apply warm-up for the first 5\% of all training steps to adjust the learning rate, a maximum input token length of 1024. We use the lora~\cite{edward2022lora} technique to fine-tune the token embedding layer and linear head layer of LLaMA2-7b for 1 epochs using AdamW optimizer on two NVIDIA RTX A6000 GPUs with a batch size of 28, a peak learning rate of $1e^{-5}$, the lora attention dimension of 16 and the alpha parameter of 32. 

\noindent{\bf For LLM inferencing}, beam search is utilized to generate a list of potential next items, evaluated under the all-item setting. To prevent the generation of non-existent IDs, we apply a constrained decoding method~\cite{wenyue2023}, setting the generation probability of invalid IDs to zero.

\begin{table}[t]
  \caption{\small Performance comparison of different methods on rating, explanation, and review task.}
  \label{tab:rating-app}
  \setlength{\tabcolsep}{1.2mm}
  \small
  \begin{adjustbox}{width=\linewidth}
  \begin{tabular}{lccccccccccccc}
    \toprule
    \multirow{2}{*}{Task Type} & \multirow{2}{*}{Metric} & \multicolumn{4}{c}{\textbf{Sports}} & \multicolumn{4}{c}{\textbf{Beauty}} & \multicolumn{4}{c}{\textbf{Toys}} \\
    \cmidrule(lr){3-6} \cmidrule(lr){7-10} \cmidrule(lr){11-14}
     & & RID & SID & CID & \textbf{META ID} & RID & SID & CID & \textbf{META ID} & RID & SID& CID & \textbf{META ID} \\
    \midrule
    \multirow{1}{*}{Rating} & RMSE 
        & \underline{1.0382} & 1.0486 & 1.0383 & \textbf{1.0327} 
        & 1.2829 & 1.3098 & \underline{1.2819} & \textbf{1.2818} 
        & \underline{1.0725} & \textbf{1.0693} & 1.0766 & 1.0770\\
    \midrule
    \multirow{2}{*}{Explan.} & BLEU-1 
        & 16.2567 & 16.5825 & \underline{16.6121} & \textbf{16.9005} 
        & 18.2299 & 18.3981 & \underline{19.3499} & \textbf{19.5106} 
        & 19.9858 & 20.2198 & \textbf{20.4570} & \underline{20.2270}\\
                                 & BLEU-4 
        & 2.1782 & 2.1944 & \underline{2.2332} & \textbf{2.3481} 
        & 2.9027 & 2.8071 & \textbf{3.0626} & \underline{3.0592} 
        & 4.3495 & 4.3701 & \textbf{4.5844} & \underline{4.4945}\\
    \midrule
    \multirow{2}{*}{Review} & BLEU-1 
        & 7.6140 & \textbf{7.7948} & 7.6586 & \underline{7.8819} 
        & 6.2282 & 6.5055 & \underline{6.5854} & \textbf{7.0500} 
        & \textbf{8.5336} & 8.0862 & 7.3846 & \underline{8.3080}\\
                            & BLEU-4 
        & 2.3228 & 1.2406 & \underline{2.4109} & \textbf{2.6546} 
        & \underline{1.9891} & 1.2406 & 1.9718 & \textbf{2.7485} 
        & 1.1315 & \textbf{1.8366} & 1.2128 & \underline{1.7061}\\
    \bottomrule
  \end{tabular}
  \end{adjustbox}
\end{table}

\subsubsection{Evaluation on Rating, Explanation and Review Tasks}

To validate META ID’s adaptability, we extend our evaluation to rating prediction, explanation generation, and review tasks, akin to P5. We use the same prompts in P5~\cite{GengRecommendation22} to convert all information into training texts. 

\noindent{\bf For LLM fine-tuning}, we pre-train T5 for 10 epochs using AdamW optimizer on two NVIDIA RTX 3090 GPUs with a batch size of 32, a peak learning rate of $1e^{-3}$. We apply warm-up for the first 5\% of all training steps to adjust the learning rate, a maximum input token length of 512 and maximum generation length of 64. 

\noindent{\bf For LLM inferencing}, greedy decoding is applied for rating prediction, explanation generation, and review tasks.

The full results are shown in Table~\ref{tab:rating-app}.

\section{Additional Experimental Results}
\label{appendix4}

\begin{figure*}[t]
\centering
  \includegraphics[width=\linewidth]{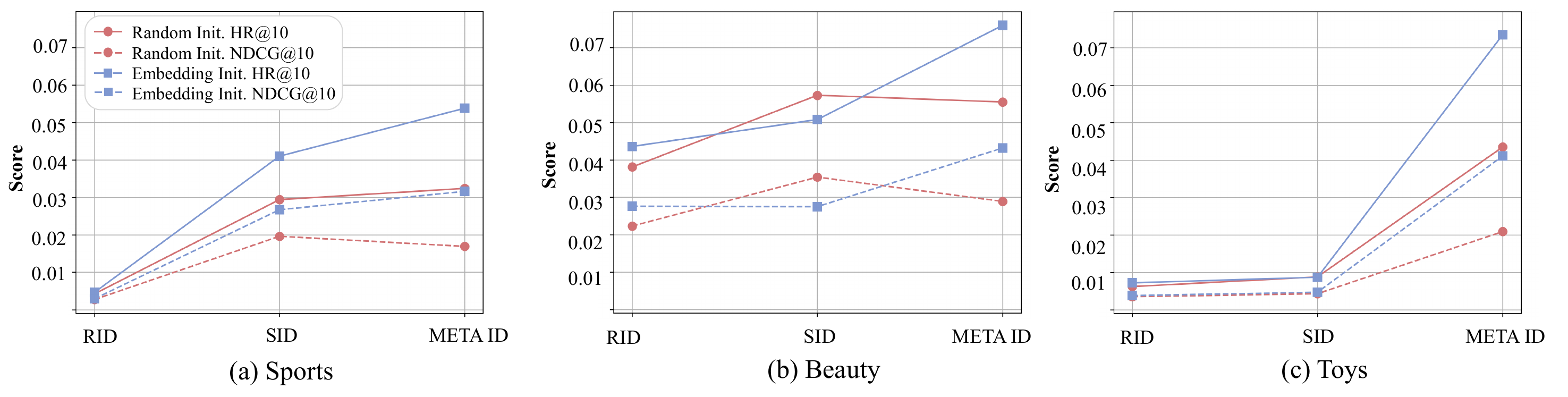}
  \caption{Performance comparision between random initialization (Random Init.) and initializing ID token embeddings using pre-trained ones (Embedding Init.). For RID and SID, random initialization does not have a positive effect compared to keeping T5’s original token embeddings. For META ID, initializing using a learned strategy with the help of cluster centroid embeddings (Embedding Init.) learned from meta-paths can significantly enhance the recommendation performance of META ID compared to random initialization.}
  
  % \Description{}
  \label{fig:init}
\end{figure*}

\noindent{\bf Initialization Approaches}. We explore the impact of different token initialization methods on the performance of META ID. Recognizing that the LLM's vocabulary includes numeric tokens for linguistic IDs, we first consider whether reinitializing these numeric tokens helps LLM for recommendation. As shown in Figure \ref{fig:init}, our experiment, contrasting random initializing numeric tokens (Random Init.) with keeping T5's original token embeddings (Embedding Init.), reveals that random initialization does not enhance performance, and be detrimental on Sports and Toys datasets. This suggests that the influence of pre-training on these tokens cannot be effectively negated through simple random initialization, thus not suitable for building IDs. This result emphasizes the importance of introducing extra tokens for IDs in LLM-based recommender systems. We also compare two initialization approaches for META ID: random initialization (Random Init.) and initializing OOV tokens using the augmentation (Embedding Init.) (See Section \ref{Section5-2}). Our findings show that the latter method substantially improves META ID's performance, underlining the critical nature of the token initialization method in achieving better results.

\begin{figure}[t!]
\centering
  \includegraphics[width=\linewidth]{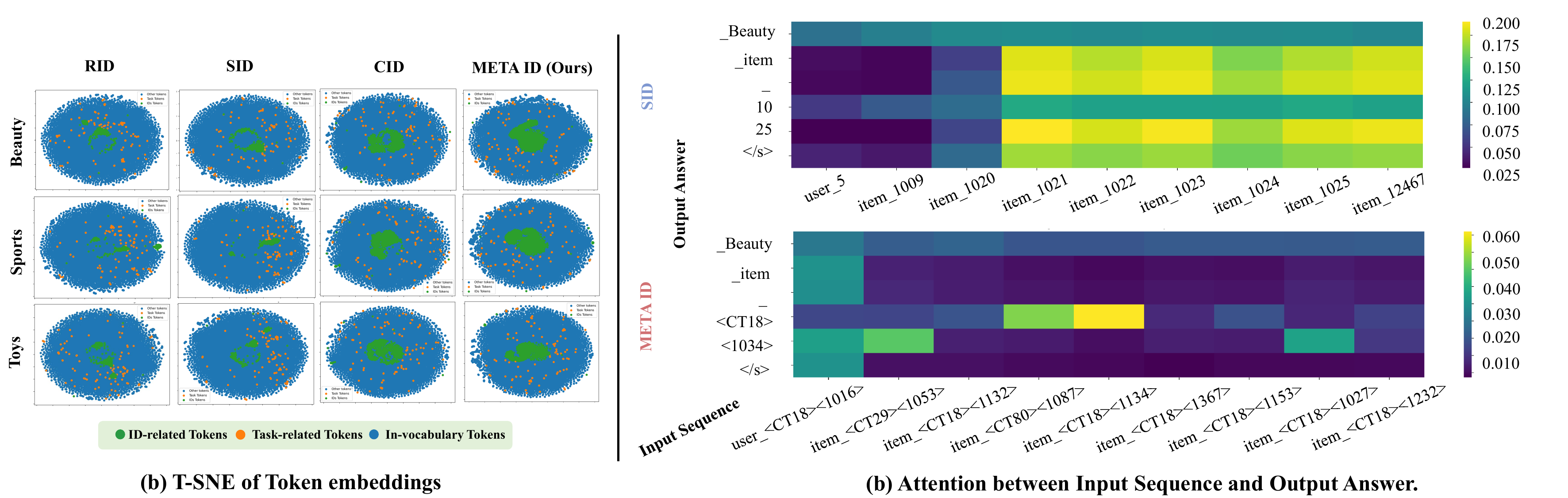}
  \caption{(a)Token embeddings projected into 2D via t-SNE, with colors indicating different type of tokens.(b) Visualization of attention between input and output tokens in sequence recommendation generation, where we aggregate tokens within one ID for clarity. The visualization indicates that SID leads to undifferentiated attention across items and neglects the user ID, whereas META ID distinctly allocates attention to sequence items, outputting item containing the "\textit{<CT18>}" token in both the user ID and item ID with the highest attention values.}
  % \Description{}
  \label{fig:tsne}
\end{figure}

\noindent{\bf Visualization of ID-related tokens}. Directly applying in-vocabulary tokens to construct IDs (RID and SID) brings poor performance in Toys dataset. In Figure~\ref{fig:tsne}a, We use t-SNE visualization to map ID token embeddings and observed that these tokens tend to be homogeneous, whereas CID and META IDs that use OOV tokens to construct IDs have a wider distribution, reflecting difference and diversity between their representations.

To further illustrate the impact on representation, we visualized the attention mechanism in sequence recommendation generation in Figure~\ref{fig:tsne}(b). This revealed that SID leads to uniform attention patterns, not distinguishing between different items and user IDs. In contrast, META ID demonstrates distinct attention patterns, successfully differentiating items and emphasizing user IDs, thereby allowing models to grasp more personalized and distinct information.

\begin{table}[t]
  \caption{The standard error of the metrics for META ID on sequential recommendation task.}
  \label{tab:br}
  \setlength{\tabcolsep}{1.6mm} % 调整列间距
  \small % 字体变小
  \begin{adjustbox}{width=\linewidth}
  \begin{tabular}{lcccc} % l for left-aligned, c for center-aligned
    \toprule
    \textbf{Datasets} & \textbf{HR@5} & \textbf{NCDG@5} & \textbf{HR@10} & \textbf{NCDG@10} \\
    \midrule
    Beauty  & 0.0510 ± 0.00038 & 0.0351 ± 0.00044 & 0.0752 ± 0.00131 & 0.0429 ± 0.00075 \\
    Sports  & 0.0322 ± 0.00061 & 0.0223 ± 0.00060 & 0.0487 ± 0.00073 & 0.0277 ± 0.00055 \\
    Toys    & 0.0503 ± 0.00091 & 0.0352 ± 0.00067 & 0.0742 ± 0.00138 & 0.0429 ± 0.00078 \\
    \bottomrule
  \end{tabular}
  \end{adjustbox}
\end{table}

\begin{table}[ht!]
\centering
\small
\caption{Average inference time (in milliseconds) of T5 with META ID on different tasks on the Toy dataset.}
\label{tab:inference-time}
\begin{tabular}{
  lcccccc
}
\toprule
\textbf{Models} & \multicolumn{2}{c}{\textbf{per user}} & \multicolumn{2}{c}{\textbf{per user-item pair}} & \multicolumn{2}{c}{\textbf{per review}} \\
\cmidrule(lr){2-3} \cmidrule(lr){4-5} \cmidrule(lr){6-7}
& {\textbf{Sequential}} & {\textbf{Direct}} & {\textbf{Rating}} & {\textbf{Explanation}} & {\textbf{Summarization}} & {\textbf{Preference}} \\
\midrule
\textbf{META ID (T)} & 74.05 & 68.60 & 5.21 & 17.28 & 9.67 & 8.55 \\
\bottomrule
\end{tabular}
\end{table}

\noindent{\bf Statistics on training \& Inference Time}. We provide statistics on the training and inference time of P5 models, we collect the running time on the Toys dataset.
As mentioned in subsection~\ref{appendix3}, we trained and test our models on two RTX 3090 GPUs. For training on sequential recommendation and direct recommendation tasks, the T5 model spent 3.5 hours to finish training. The average inference time of T5 model on dferent tasks are presented in Table~\ref{tab:inference-time}. Sequential and direct recommendation tasks require much longer inference time than other tasks due to the beam search step. Overall, the inference is very fast. It is also promising to further reduce the training and inference time with the help of effcient Transformer techniques.

\section{Discussions}
\label{appendix5}

There are two promising directions of META ID. First, META ID uses a fixed database of users and items, while newly appearing items and users do not have interaction history. This could be solved using methods related to the cold start issue. Second, META ID applies two-level tokens for constructing IDs, while a more complicated hierarchical structure could be considered. Then, META ID could be applied to modern recommender system containing trillions of users and items.

%%%%%%%%%%%%%%%%%%%%%%%%%%%%%%%%%%%%%%%%%%%%%%%%%%%%%%%%%%%%